\documentclass[11pt]{article} 

\usepackage[utf8]{inputenc} 

\usepackage[margin=1in]{geometry}
\usepackage{graphicx} 

\usepackage{booktabs} 
\usepackage{array} 
\usepackage{paralist} 
\usepackage{verbatim} 
\usepackage{subfig} 
\usepackage{xcolor} 

\usepackage{fancyhdr} 
\usepackage{sectsty}
\allsectionsfont{\sffamily\mdseries\upshape} 

\usepackage[nottoc,notlof,notlot]{tocbibind} 
\usepackage[titles,subfigure]{tocloft} 

\usepackage{amsmath,amsthm}
\usepackage{amssymb}

\newtheorem{theorem}{Theorem}
\newtheorem{lemma}{Lemma}
\newtheorem{corollary}{Corollary}
\newtheorem{definition}{Definition}
\newtheorem{conjecture}{Question}
\newtheorem{axiom}{Assumption}

\newtheorem{convention}{Convention}

\DeclareMathOperator{\tr}{tr}
\DeclareMathOperator{\Hess}{Hess}

\title{Flatly Foliated Relativity}
\author{Hubert L. Bray\footnote{H. Bray acknowledges the support of NSF Grant DMS-1406396.}, Benjamin Hamm, Sven Hirsch, James Wheeler and Yiyue Zhang}

\date{\today} 

\begin{document}

\maketitle

\begin{abstract}
Flatly Foliated Relativity (FFR) is a new theory which conceptually lies between Special Relativity (SR) and General Relativity (GR), in which spacetime is foliated by flat Euclidean spaces.  While GR is based on the idea that ``matter curves spacetime", FFR is based on the idea that  ``matter curves spacetime, but not space''.  This idea, inspired by the observed spatial flatness of our local universe, is realized by considering the same action as used in GR, but restricting it only to metrics which are foliated by flat spatial slices. FFR can be thought of as describing gravity without gravitational waves.

In FFR, a positive cosmological constant implies several interesting properties which do not follow in GR: the metric equations are elliptic on each euclidean slice, there exists a unique vacuum solution among those spherically symmetric at infinity, and there exists a geometric way to define the arrow of time. Furthermore, as gravitational waves do not exist in FFR, there are simple analogs to the positive mass theorem and Penrose-type inequalities. 

Importantly, given that gravitational waves have a negligible effect on the curvature of spacetime, and that the universe appears to be locally flat, FFR may be a good approximation of GR.  Moreover, FFR still admits many notable features of GR including the big bang, an accelerating expansion of the universe, and the Schwarzschild spacetime. Lastly, FFR is already known to have an existence theory for some simplified cases, which provokes an interesting discussion regarding the possibility of a more general existence theory, which may be relevant to understanding existence of solutions to GR.

\end{abstract}

\section{Introduction}

It is an honor and a privilege to contribute an article to this volume 
celebrating the 60th birthday of Robert Bartnik. Robert is one of the 
most influential researchers who have used geometry and analysis to 
understand deep questions in general relativity. From showing that the 
total ADM mass of a spacetime is well-defined, to his formulation of the 
Bartnik mass, it is impossible to do geometric relativity without 
encountering many of Robert's wonderful ideas. His friendship, advice, 
and encouragement for young people has also helped make the geometric 
analysis community what it is today.

In this paper, we define {\it flatly foliated relativity} (FFR) to be a new theory of gravity in which the spatial flatness of the universe is strictly required.  This requirement is captured in the following fundamental assumption:

\begin{axiom}\label{axiom:flat}
Spacetime is foliated by flat three dimensional Euclidean spaces.
\end{axiom}

The purpose of this paper is to investigate the consequences of this assumption while otherwise leaving the action of GR unchanged. Recall that this action contains a parameter $\Lambda$, referred to as the \textit{cosmological constant}; we will see that several qualitative features of FFR depend on the sign of $\Lambda$.

On the one hand, FFR is general enough to include several of the most important spacetimes of GR such as the Schwarzschild solution and the Friedmann–Lemaître–Robertson–Walker (FLRW) spacetimes with spatial curvature $k=0$. More generally, any spherically symmetric spacetime can be flatly foliated if there exists a single flat slice \cite{guven1999flat}. There are also analogues to the positive mass theorem \cite{PMT1}, \cite{PMT2} and Penrose inequality \cite{penrose1}, \cite{penrose2}.

On the other hand, FFR is simpler than GR in the sense that, when the cosmological constant is positive, the main equations are elliptic rather than hyperbolic. In particular, this has the consequence that we obtain an existence theory when the vector field $J$, the current density of energy and momentum as measured by an observer moving perpendicular to the flat foliation, vanishes. In comparison, the existence theory for GR - even in special cases - is still a big open problem.

\vspace{.1in}
\noindent
Some similarities to GR include:
\begin{enumerate}
\item The universe is locally modeled by SR. In particular there is a special speed, called the speed of light, which is locally the same for all observers.
\item Matter which may not go faster than the speed of light in GR also may not go faster than the speed of light in FFR.
\item The big bang is modeled by FLRW spacetimes, but with $k = 0$.
\item Spacetime outside black holes, gravitational lensing, the gravitational redshift of light, and the gravity of spherically symmetric bodies like stars and planets, including the precession of the perihelion of the orbit of Mercury, may be modeled by the exterior region of the Schwarzschild spacetime, which is foliated by flat slices.

\end{enumerate}
Some differences to GR include:
\begin{enumerate}
\item {A positive cosmological constant implies the existence of a well-defined geometric arrow of time, whereas a negative or zero cosmological constant does not.}

\item A positive cosmological constant implies that there exists a unique vacuum solution among solutions which are spherically symmetric at infinity, whereas a negative or zero cosmological constant does not. In fact, we will prove a result stronger than this in section \ref{sec:vacuum}.

\item A positive cosmological constant makes the resulting spacetime metric equations elliptic on each Euclidean space, a highly desirable analytic feature when trying to prove existence, uniqueness, and regularity of solutions, whereas a negative or zero cosmological constant does not.

\item Gravitational waves moving at the speed of light do not exist. 

\item The speed of gravity is infinity, similar to Newtonian physics.

\end{enumerate}

As seen in the list above, FFR has several nice properties when the cosmological constant is positive, and further these properties are not generically present otherwise. For this reason, we propose another assumption on FFR spacetimes:

\begin{axiom}\label{axiom:arrow}
The cosmological constant $\Lambda$ is strictly positive.
\end{axiom}

In addition to the nice properties it yields, this assumption is physically motivated by observations of the universe which indicate a positive cosmological constant within the standard $\Lambda$CDM model of cosmology \cite{AcceleratingUniverse2}, \cite{AcceleratingUniverse1}. The first of the nice properties we've mentioned is that assumption \ref{axiom:arrow} gives rise to a well-defined, geometrically preferred arrow of time.

An arrow of time separates the notion of the future from the past. There are two connected components of time-like vectors at every point of a spacetime; an arrow of time defines, globally and in a continuous manner, which component represents future time-like vectors. For spatially foliated spacetimes like ours, an arrow of time is equivalent to a globally and continuously defined unit time-like vector perpendicular to the space-like foliation. We will show in section \ref{sec:arrow} that assumption \ref{axiom:arrow}, along with a nonnegative energy density, allows us to choose the direction of the flat foliation's mean curvature vector as our arrow of time. Further, with this choice, the volume form of the flat foliation is always getting bigger, in agreement with the cosmological arrow of time given by the universe's expansion in the FLRW spacetime.

Our main result is summarized as follows:

\begin{theorem}\label{thm:intro}
Suppose assumption \ref{axiom:arrow} holds and consider the FFR spacetimes with non-negative energy density at a critical point of the Einstein Hilbert action when varied with respect to spacetimes satisfying assumption \ref{axiom:flat}.
Then

\begin{itemize}
\item  For each such spacetime, there exists a geometrically preferred arrow of time.
\item For each such spacetime, the volume form of the flat foliation is always getting bigger with respect to this arrow of time. 
\item  For each such spacetime, the equations describing the evolution of the spacetime metric are elliptic on each Euclidean slice, and uniformly elliptic assuming the bounds in equation \ref{eqn:precisebounds}.
\item There exists a unique vacuum solution among solutions which are spherically symmetric at infinity.
\end{itemize}
\end{theorem}
The fact that such innocent looking assumptions can have such nontrivial consequences makes these assumptions very interesting to study. 
These assumptions are also compatible with the axioms proposed by the author in \cite{bray2010dark}, which imply the Einstein-Hilbert action with a cosmological constant, as well as a geometrically natural model for dark matter. However, assumptions \ref{axiom:flat} and \ref{axiom:arrow} are logically independent from those axioms, so we will focus only on the former in this paper.

\vspace{.1in}
\noindent
While FFR is not a candidate for the true theory of gravity due to the observation of gravitational waves \cite{LIGO}, there are many reasons that studying FFR is useful:
\begin{enumerate}
\item Approximating GR in computer simulations. The equations of FFR are simpler to solve than those of GR in many ways, in part because gravitational waves do not exist in the theory. There is also a canonical foliation.
\item Exact solutions for FFR may be interpreted as approximate solutions for GR.
\item Understanding the theoretical properties of GR. Take your favorite open problem in GR, and then solve it first for FFR, which may be quite a bit easier. 
\item As seen in section \ref{sec:Schwarzschild}, FFR avoids the insides of black holes in the Schwarzschild case and at least somewhat more generally, which may be important for problems where avoiding spacetime singularities is useful.
\item Applications of elliptic techniques. Whereas the Einstein equation for GR is hyperbolic, the analogous equations for FFR are elliptic.
\end{enumerate}

A large hurdle in understanding the existence theory of GR is understanding how to handle the singularities which arise inside black holes. This problem is actually avoided in FFR, as stated in item 4 above. This, along with the fact that FFR is already known to have an existence theory for special cases ($J$ = 0), suggests that it could be possible for FFR to have a more general existence theory. If this is true, then another question arises: how could such an existence theory of FFR be used to understand existence of solutions in GR? Though it is first necessary to understand whether or not FFR has a general existence theory, this question regarding GR provides relevance to approaching the problem in FFR.

Our paper is structured as follows:
in sections 2 and 3 we establish geometric preliminaries and derive the fundamental FFR equations.
These will be studied in more detail in section 4 and 5 where we prove in particular that  assumption \ref{axiom:arrow} implies the existence of a geometrically preferred arrow of time, the adoption of which implies ellipticity.
In section 6, we analyze some explicit examples such as the Schwarzschild spacetime and the FLRW spacetimes. 

\section{Geometric Preliminaries}

Another nice consequence of assumption \ref{axiom:flat} is that it is always possible to choose globally defined coordinates. In this section we define the spacetime metric of FFR in very natural coordinates, as well as the second fundamental form of each flat, constant time slice of the flat foliation. We also establish conventions in our notation.

\subsection{The Spacetime Metric}

We choose $(t = x^0, x = x^1, y = x^2, z = x^3)$ as our coordinates so that the spacetime metric is globally expressed as 
\begin{equation}\label{eqn:Metric}
g_{\alpha \beta} =
\begin{bmatrix}
-\eta^2 + |\omega|^2 &  -\omega_1 & -\omega_2 & -\omega_3 \\
      -\omega_1        &     1   &    0   &    0   \\
      -\omega_2        &     0   &    1   &    0   \\
      -\omega_3        &     0   &    0   &    1 
\end{bmatrix}  
= (-\eta^2 + |\omega|^2) dt^2 - (\omega \otimes dt + dt \otimes \omega) + \delta_{{\bf R}^3}
\end{equation}
where 
$\omega = \omega_1 dx^1 + \omega_2 dx^2 + \omega_3 dx^3$ is a one form, 
$|\omega|^2 = \omega_1^2 + \omega_2^2 + \omega_3^2$, and 
$\delta_{{\bf R}^3}$ is the flat metric on ${\bf R}^3$. 
Note that $\det(g_{\alpha \beta}) = -\eta^2$, so the above form can be realized for any flatly foliated spacetime with signature $-+++$. Further note that scaling the pair ($\omega,\eta$) by a positive multiplicative constant is equivalent to doing the same to the coordinate time $t$.

Equation \ref{eqn:Metric} highlights how FFR is a cross between SR and GR. When $\eta \equiv 1$ and $\omega \equiv 0$, the spacetime metric $g$ is the Minkowski spacetime metric of SR. On the other hand, GR allows for any spacetime metric, not one necessarily restricted to the form of equation \ref{eqn:Metric}. Said another way, FFR is spatially like SR, but otherwise like GR.

On the other hand, because of the coordinate chart invariance for GR, FFR is more like GR than SR, in the following sense: whereas the components of a general spacetime metric are 10 functions, 4 of these functions may be effectively specified by a choice of coordinate chart. Hence, GR is locally described by 6 functions on the spacetime, which is only 2 more than the 4 functions $\eta, \omega_1, \omega_2, \omega_3$ which describe FFR.

Define $\partial_k = \frac{\partial}{\partial x^k}$. We will raise and lower indices of three dimensional vectors and covectors using the flat metric on ${\bf R}^3$ so that, for example, $\omega^k = \omega_k$. We will tend to stick with the lowered indices. 
Because of this, whenever we repeat an index, whether raised or lowered, summation will be implied.
As usual, we will use New Roman indices to range from 1 to 3 and Greek indices to range from 0 to 3.

Thus, for example, the dual vector $\vec\omega$ to $\omega$ is just
\begin{equation}
\vec{\omega} = \omega_i \partial_i = \omega_1 \partial_1 + \omega_2 \partial_2 + \omega_3 \partial_3 .
\end{equation}
The orthogonal flow vector for the flat, constant time slices is 
\begin{equation}
   \vec\eta = \partial_t + \vec\omega = (1, \omega_1, \omega_2, \omega_3) \mbox{ in coordinates,}
\end{equation}
which is orthogonal to each constant time slice since $g(\vec\eta, \partial_k) = 0$, for $k = 1,2,3$. Also note that $g(\vec\eta,\vec\eta) = -\eta^2$, so $\eta$ is in fact the length of the time-like vector $\vec\eta$. Hence, if we define 
\begin{equation}
n = \frac{\vec\eta}{\eta}, 
\end{equation}
an orthonormal basis at each point is 
$\{n, \partial_1, \partial_2, \partial_3\}$.

We will assume that everything is smooth and that $\eta > 0$ so that the inverse of $g$ exists and is smooth too. 
Note that 
\begin{equation}\label{eqn:ginv}
g^{\alpha \beta} =
\begin{bmatrix}
-\frac{1}{\eta^2} &  -\frac{\omega_1}{\eta^2} & -\frac{\omega_2}{\eta^2} & -\frac{\omega_3}{\eta^2} \\
-\frac{\omega_1}{\eta^2} & 1-\frac{\omega_1^2}{\eta^2}         & -\frac{\omega_1\omega_2}{\eta^2} & -\frac{\omega_1\omega_3}{\eta^2} \\
-\frac{\omega_2}{\eta^2} & -\frac{\omega_2\omega_1}{\eta^2} & 1-\frac{\omega_2^2}{\eta^2}         & -\frac{\omega_2\omega_3}{\eta^2}   \\
-\frac{\omega_3}{\eta^2} & -\frac{\omega_3\omega_1}{\eta^2} & -\frac{\omega_3\omega_2}{\eta^2} & 1-\frac{\omega_3^2}{\eta^2}
\end{bmatrix}  
= - \left( \frac{\vec\eta}{\eta} \right)^2 + \delta_{{\bf R}^3}
= - n \otimes n + \delta_{{\bf R}^3}   ,
\end{equation}
which is easiest to see when $\vec\omega$ is pointing in the $x^1$ direction.

\subsection{The Second Fundamental Form}

Now we will show that the second fundamental form $\vec{k} = k n$ of each constant time slice is 
\begin{equation}\label{eqn:SFF}
   k_{ij} = \frac{\omega_{i,j} + \omega_{j,i}}{2\eta}, 
\end{equation}
so that the mean curvature of each constant time slice is 
\begin{equation}\label{eqn:mc}
   K = \tr k_{ij} = \frac{\omega_{i,i}}{\eta} = \frac{\nabla \cdot \omega}{\eta}, 
\end{equation}
where $\nabla \cdot \omega$ is the divergence of $\omega$ on ${\bf R}^3$. 

An elegant way to derive 
equation \ref{eqn:SFF} using other well known formulas is to use the fact that $\eta = \partial_t + \vec\omega$ is the orthogonal flow vector for the constant time slices. Expressing the rate of change of the flat metrics as a function of time in terms of their second fundamental forms gives
\begin{equation}
    \frac{d}{dt} \delta_{{\bf R}^3} = - 2 \langle \vec\eta, \vec{k}_{ij} \rangle = 2 \eta \, k_{ij} .
\end{equation}
Meanwhile, the rate of change of the flat metrics also just amounts to a reparametrization of these metrics, which may be expressed in terms of the Lie derivatives of the flat metrics,
\begin{equation}
    \frac{d}{dt} \delta_{{\bf R}^3} = \mathcal{L}_{\vec\omega}(\delta_{{\bf R}^3}) = \omega_{i,j} + \omega_{j,i}.
\end{equation}
Together, the last two equations imply equation \ref{eqn:SFF}.

To compute the second fundamental form directly, we need to know one of the Christoffel symbols \cite{ONeill} of the spacetime metric. For handy future reference, we include all of them here, where $1 \le i,j,k,m \le 3$:
\begin{eqnarray}
\Gamma_{00}^{\;\;\;\;0} &=& \frac{\eta_0}{\eta} - \frac{\omega_m \, \eta_m}{\eta}    
                                             + \frac{ \omega_j \, \omega_{j,m} \, \omega_m  }{\eta^2} \\
\Gamma_{00}^{\;\;\;\;k} &=& -\omega_{k,0} - \omega_m \, \omega_{m,k} + \eta \, \eta_k 
                                          + \omega_k \; \Gamma_{00}^{\;\;\;\;0} \\
\Gamma_{i0}^{\;\;\;\;0} &=& \frac{\eta_i}{\eta} - \frac{\omega_m}{2\eta^2}
                                             (\omega_{i,m} + \omega_{m,i}) \\
\Gamma_{i0}^{\;\;\;\;k} &=& \frac12 (\omega_{i,k} - \omega_{k,i}) + \omega_k \; \Gamma_{i0}^{\;\;\;\;0} \\
\Gamma_{ij}^{\;\;\;\;0} &=& \frac{1}{2\eta^2} (\omega_{i,j} + \omega_{j,i})     \\
\Gamma_{ij}^{\;\;\;\;k} &=& \omega_k \; \Gamma_{ij}^{\;\;\;\;0}
\end{eqnarray}

Since $n$ is the future pointing normal vector to each flat constant time slice, the second fundamental form may be expressed as 
\begin{equation}
   \vec{k}(\partial_i, \partial_j) = \mbox{nor}(\nabla_{\partial_i} \partial_j) = k(\partial_i, \partial_j) n,
\end{equation}
where $k$ is real-valued. Hence,
\begin{eqnarray}\nonumber
k_{ij} &=& k(\partial_i, \partial_j) = - \langle n , \nabla_{\partial_i} \partial_j \rangle
= - \langle n , \Gamma_{ij}^{\;\;\;\;0} \, \partial_t + \Gamma_{ij}^{\;\;\;\;m} \, \partial_m \rangle \\
&=& - \left\langle \frac{\partial_t + \vec{\omega}}{\eta}  \; , \; \Gamma_{ij}^{\;\;\;\;0} \, (\partial_t + \vec{\omega}) \right\rangle 
= \eta \; \Gamma_{ij}^{\;\;\;\;0}
= \frac{1}{2\eta} (\omega_{i,j} + \omega_{j,i})  ,
\end{eqnarray}
where to go from the first line to the second line, we used the fact that $n$ is perpendicular to all of the vectors tangent to the flat slice.

As a final comment, the second fundamental form $k_{ij}$ is a geometric invariant and hence may be thought of as a physical quantity. By contrast, $\omega$ is not a geometric invariant at any given point and, with a change of coordinates, may even be chosen to be zero along any smooth time-like curve. Since $\eta$ is the orthogonal flow speed of the flat slices, it is a physical quantity, but only up to multiplying by a function of $t$ (corresponding to the freedom in how one parametrizes the flat slices in $t$).

\section{Derivation of the Fundamental Equations}

We begin with a quick review of how the Einstein equation is derived from the Einstein-Hilbert action of GR. We then adapt these computations to FFR.

\subsection{A Quick Review of General Relativity}

General Relativity is most precisely defined by saying that the universe (including the spacetime metric and all of its matter fields) is at a critical point of a real-valued function $\mathcal{F}$ called the action. The Einstein-Hilbert action of GR, where the speed of light and the universal gravitational constant have been set to one, is 
\begin{equation}\label{eqn:action}
   \mathcal{F} = \int (S - 2\Lambda + 16\pi L) \; dV,
\end{equation}
where $S$ is the scalar curvature of the spacetime, $dV$ is the volume form, $\Lambda$ is a fixed constant of nature called the cosmological constant, and $L$ represents additional matter field terms which may be added.
As one might guess, the $-2$ and the $16\pi$ are completely arbitrary conventions, often chosen to reproduce the usual form of the Einstein equation. Though the cosmological constant is occasionally considered part of $L$ in other treatments, we choose to keep them separate.
We will also assume, as is standard, that $L$ may be expressed geometrically so that the action is invariant under reparametrizations of the metric. 

For example, to derive the Einstein-Maxwell equations, let  
$L = -\frac14 |dA|_g^2$, where $A$ is the co-vector potential of electromagetism.
(Typically, $L$ involves the spacetime metric $g$ as well as the matter fields: in this case, the metric $g$ is used to take the norm of $dA$.)
The spacetime metric $g$ and the one form $A$ would then both have to be chosen to be at a critical point of $\mathcal{F}$. These critical point equations are called the Euler-Lagrange equations for $g$ and $A$. The Euler-Lagrange equation for $A$ is the Maxwell equation $d^* dA = 0$ on a spacetime which describes how $A$ (and hence the Faraday tensor $F = dA$ whose components are the electric and magnetic fields) evolves over time.

In this paper, we will focus on the Euler-Lagrange equation for $g$, which as we will show is the Einstein equation with a cosmological constant. Note that the integral in the action is taken over smooth open bounded regions of the spacetime and that $g$ will be required to be at a critical point for all of these regions, for all variations of $g$ compactly supported inside the region.

Given a one parameter family of spacetime metrics $g(s)$, let $\dot{g} = \frac{d}{ds}g(s)|_{s=0}$. Continuing with this dot notation, standard calculations give us that
\begin{eqnarray}
   \dot{S} &=& -\langle Ric, \dot{g} \rangle + \nabla \cdot (\nabla \cdot \dot{g}) - \Delta (\tr g) \\
              &=& -\langle Ric, \dot{g} \rangle + \nabla \cdot ( \mbox{irrelevant stuff})
\end{eqnarray}
Combining this with $\dot{dV} = \frac12 \langle g, \dot{g} \rangle \; dV$ and using the divergence theorem to get rid of the irrelevant divergence terms, we compute the standard formula for the first variation $\dot{\mathcal{F}}$ of the action $\mathcal{F}$:
\begin{equation}\label{eqn:FV}
\dot{\mathcal{F}} = \int \langle \dot{g} \; , \; 8\pi T - \Lambda g - G \rangle \; dV
\end{equation}
where $G = Ric - \frac12 S g$ is the Einstein curvature tensor and $T$, called the stress-energy tensor, collects all of the terms coming from $L$ by definition. Note that the divergence theorem boundary term is zero since $\dot{g}$ is zero there.

The only way for $\dot{\mathcal{F}}$ to equal zero for all possible $\dot{g}$ is for 
\begin{equation}
   G = 8\pi T - \Lambda g,
\end{equation}
which is the Einstein equation with a cosmological constant. Note that the cosmological constant is often absorbed into the definition of $T$. 

A \textit{solution} to GR, then, is a spacetime manifold equipped with a smooth metric $g$ and associated matter fields satisfying their equations of motion such that $ G=8 \pi T-\Lambda g$.

\subsection{Flatly Foliated Relativity}\label{sec:FFR}

Equation \ref{eqn:FV} is still true for FFR. However, because the metric $g$ has a restricted form due to equation \ref{eqn:Metric}, the variation of the action $\dot{\mathcal{F}}$ does not have to equal zero for all $\dot{g}$. Instead, we must differentiate equation \ref{eqn:Metric},
\begin{equation}
   \dot{g} = (-2 \eta \dot{\eta} + 2\langle \omega, \dot{\omega} \rangle) dt^2
                 - (\dot\omega \otimes dt + dt \otimes \dot\omega) ,
\end{equation}
and then plug this in to equation \ref{eqn:FV}, to get
\begin{eqnarray}
\dot{\mathcal{F}} 
&=& \int g^{\alpha\gamma} \; g^{\beta\theta} \; \dot{g}_{\alpha\beta} \; (8\pi T - \Lambda g - G)_{\gamma\theta} \\
&=& \int (-n^\alpha n^\gamma + \delta_{{\bf R}^3}^{\alpha\gamma}) \; (-n^\beta n^\theta + \delta_{{\bf R}^3}^{\beta\theta}) \; (8\pi T - \Lambda g - G)_{\gamma\theta}  \\ \nonumber
&& \;\;\;\; \left[ (-2 \eta \dot{\eta} + 2\langle \omega, \dot{\omega} \rangle ) 
\; \delta_{0\alpha} \; \delta_{0\beta} - \dot\omega_\alpha \; \delta_{0\beta} 
- \dot\omega_\beta \; \delta_{0\alpha} \right] \; dV , \\
&=& 2 \int \left[ -\dot{\eta} \;  (8\pi T - \Lambda g - G)(n,n) + \dot{\omega}_i  (8\pi T - \Lambda g - G)(n,\partial_i) \right]
 \,dt\,dx\,dy\,dz
\end{eqnarray}
where we used equation \ref{eqn:ginv} to go from the first line to the second line and the fact that 
\begin{equation}
dV = |\det(g_{ij})|^{1/2} \,dt\,dx\,dy\,dz = \eta \,dt\,dx\,dy\,dz 
\end{equation}
to go from the second line to the third, which also involves a fair amount of algebra.

The only way for $\dot{\mathcal{F}}$ to equal zero for all possible $\dot{\eta}$ and $\dot{\omega}_i$ is for
\begin{eqnarray}
0 &=& (8\pi T - \Lambda g - G)(n,n) \label{eqn:Enn} \\ \label{eqn:Enp}
0 &=& (8\pi T - \Lambda g - G)(n,\partial_i), \mbox{ for } i = 1,2,3.
\end{eqnarray}
In other words, FFR only implies that the Einstein equation must be satisfied in the direction of $n$. Said another way, only the energy and momentum density as observed by an observer going in the direction $n$, the direction perpendicular to the flat foliation, affects the curvature of the spacetime. 

A \textit{solution} to FFR, then, is a spacetime manifold equipped with a smooth metric $g$, associated matter fields satisfying their equations of motion, and a particular flat foliation by Euclidean spaces such that equations \ref{eqn:Enn} and \ref{eqn:Enp} are satisfied. Again, this is not as far from GR as one might think. While at first glance it appears that we are ignoring 6 components of the stress energy tensor, the fact that the stress energy tensor has zero divergence (which is a vector equation) means we are actually only ignoring 2 degrees of freedom of the stress energy tensor. That is, FFR curves spacetime using 4 of the 6 degrees of freedom of the stress energy tensor. Note that a solution to FFR need not still be a solution if one considers a different flat foliation; however, given a solution to GR, it is a solution to FFR for any choice of flat foliation, assuming one exists.

Now we will express our fundamental equations in terms of $\eta$ and $\omega$. The Gauss and Codazzi equations imply that 
\begin{eqnarray}
   G(n,n) &=& \frac12 ((\tr k)^2 - ||k||^2) \\
   G(n,\partial_i) &=& (\nabla \cdot k)_i - d(\tr k)_i
\end{eqnarray}
since the scalar curvature of each constant time slice is zero. Also, as is standard, define
\begin{eqnarray}
  \mu &=& T(n,n) \\
  J_i &=& T(n,\partial_i)
\end{eqnarray}
Plugging in equation \ref{eqn:SFF} for $k$ and simplifying, we get

\vspace{.1in}
\begin{center}
{\bf The Flatly Foliated Relativity Equations} 
\end{center}
\begin{eqnarray}
   2\Lambda + 16\pi \mu  &=& \frac{1}{\eta^2}\left[(\nabla \cdot \omega)^2 - \frac14 ||\omega_{i,j} + \omega_{j,i}||^2 \right]  
\label{eqn:NO1}  \\  \label{eqn:NO2}
   16\pi J_i &=& \frac{1}{\eta}\left( \Delta\omega_i - d (\nabla \cdot \omega)_i  \right)
                     + \frac{2}{\eta^2}(\nabla \cdot \omega) \eta_i 
                     - \frac{1}{\eta^2}(\omega_{i,j} + \omega_{j,i})\eta_j 
\end{eqnarray}
where we are using the standard convention that $\omega_{i,j}$ is the coordinate derivative of $\omega_i$ in the direction of $\partial_j$, $\nabla \cdot \omega = \omega_{i,i}$ (summation of $i$ from $1$ to $3$ implied) is the divergence of $\omega$, $\Delta \omega_i = \omega_{i,jj}$ is the Laplacian of the flat metric on ${\bf R}^3$ of $\omega_i$,
$d (\nabla \cdot \omega)_i = \omega_{j,ji}$, and 
$||\omega_{i,j} + \omega_{j,i}||^2 = \sum_{i,j = 1}^3 (\omega_{i,j} + \omega_{j,i})^2$.
Note that since each constant time slice is flat, the coordinate derivatives are equal to the covariant derivatives. Also, instead of writing $\eta_{,i}$ to denotes the derivative of $\eta$ in the direction of $\partial_i$, we will just write $\eta_i$. 

One feature of the above equations which is remarkable is that there are not any time derivatives! That is, it appears that $\eta$ and $\omega$, and hence the metric on each constant time slice, are determined by equations only involving spatial derivatives on each slice. Since the metric determines the paths of geodesics which have gravitational interpretations, this fact is reminiscent of the Poisson equation for Newtonian gravity.

But how does one solve for $\eta$ and $\omega$ given $\mu$ and $J$? To make progress on this question, we note that equation \ref{eqn:NO2} implies the useful and somewhat surprising identity
\begin{equation}\label{eqn:NO3}
   8\pi \; \nabla \cdot (J\,\eta^2) = (\nabla \cdot \omega) \Delta \eta - \frac12 (\omega_{i,j} + \omega_{j,i}) \eta_{ij}.
\end{equation}
Even more progress is made by replacing $\omega$ with two new variables, $f$ and $\zeta$, where we let 
\begin{equation}\label{eqn:new}
   \omega = df + \zeta
\end{equation}
for some real-valued function $f$ and one form $\zeta$, chosen so that 
\begin{equation}
\nabla \cdot \zeta = 0.
\end{equation}
This decomposition, which is equivalent to solving $\Delta f = \nabla \cdot \omega$, can often be achieved. For example, the Helmholtz decomposition theorem yields such a decomposition if $\omega$ decays faster than $1/|x|$ \cite{griffiths2005introduction}, and in the spherically symmetric case, where $\omega = \omega(r)dr$, this is solved with $\zeta \equiv 0$ by integrating $f'(r)=\omega(r)$.
Beautifully, when we plug equation \ref{eqn:new} into equations \ref{eqn:NO1}, \ref{eqn:NO2}, and \ref{eqn:NO3},
the third derivatives of $f$ cancel out and the equation for $\zeta$ simplifies nicely, resulting in the following system of partial differential equations (PDEs) for $f$, $\zeta$, and $\eta$:

\vspace{.1in}
\begin{center}
{\bf The Flatly Foliated Relativity Equations in Elliptic Form} 
\end{center}
\begin{eqnarray}
\label{eqn:FFR1} (\Delta f)^2 - ||Q_{ij}||^2 &=& (16\pi \mu + 2\Lambda) \; \eta^2 \\
\label{eqn:FFR2} \Delta \zeta_i &=& 16\pi J_i \; \eta - \frac{2}{\eta}\left( \Delta f \cdot \delta_{ij} - Q_{ij} \right) \eta_j \\
\label{eqn:FFR3} \left( \Delta f \cdot \delta_{ij} - Q_{ij} \right) \eta_{ij} &=& 8\pi \nabla \cdot (J\,\eta^2)
\end{eqnarray}
\noindent
where $Q_{ij} = f_{ij} + \frac12 (\zeta_{i,j} + \zeta_{j,i})$ and we require that $\lim_{x\rightarrow \infty} \nabla \cdot \zeta = 0$ (which theorem \ref{thm:equivalence} shows implies $\nabla \cdot \zeta = 0$ everywhere).  Equivalently, all three equations may be written in divergence form:
\begin{eqnarray}
\label{eqn:FFR1D}
\nabla _j \cdot \left( f_{ii} \, f_j - f_{ij}  \, f_i - 2 f_i \, \zeta_{j,i}\right) 
&=& \frac14 ||\zeta_{i,j} + \zeta_{j,i}||^2 + (16\pi \mu + 2\Lambda) \; \eta^2 \\
\label{eqn:FFR2D}
\nabla _j \cdot \left(  \zeta_{i,j} \right) &=& 16\pi J_i \; \eta - \frac{2}{\eta}\left( \Delta f \cdot \delta_{ij} - Q_{ij} \right) \eta_j \\
\label{eqn:FFR3D}
\nabla _j \cdot \left( f_{ii} \,  \eta_j - f_{ij}  \, \eta_i - \zeta_{j,i}  \, \eta_i \right) &=& 8\pi \nabla \cdot (J\,\eta^2)
\end{eqnarray}
When $\mu \ge 0$ and $\Lambda > 0$, the above system of equations is elliptic, as shown in section \ref{sec:elliptic}.

\vspace{.25in}
{\it Comment: } The fact that the above system of equations may be written in divergence form allows for weak solutions in the distributional sense to be defined. For example, in section \ref{sec:SchwarzschildSolution} the Schwarzschild spacetime is realized as the solution to the above system of equations when $\mu = m \, \delta_0$,  $J = 0$, and $\Lambda = 0$, where $\delta_0$ is the Dirac-delta distribution at the origin.

\vspace{.1in}
Equations \ref{eqn:NO1} and \ref{eqn:NO2} are a system of PDEs in $(\eta,\omega)$, whereas equations \ref{eqn:FFR1}, \ref{eqn:FFR2}, and \ref{eqn:FFR3} are a system of PDEs in $(\eta, f, \zeta)$. It turns out these systems are equivalent under very general circumstances.
\begin{theorem} \label{thm:equivalence} For smooth solutions with $\eta > 0$, \\
equations \ref{eqn:NO1} and \ref{eqn:NO2} plus the requirement that there exists an $f$ which solves 
$\Delta f = \nabla \cdot \omega$
are equivalent to 
equations \ref{eqn:FFR1}, \ref{eqn:FFR2}, and \ref{eqn:FFR3} plus the requirement that $\lim_{x\rightarrow \infty} \nabla \cdot \zeta = 0$.
\end{theorem}
{\it Proof: } Suppose we are given $\eta$ and $\omega$ which solve equations \ref{eqn:NO1} and \ref{eqn:NO2} and an $f$ which solves $\Delta f = \nabla \cdot \omega$. Let $\zeta = \omega - df$ and note that $\nabla \cdot \zeta = 0$. Substituting $\omega = df + \zeta$ into equations \ref{eqn:NO1}, \ref{eqn:NO2}, and \ref{eqn:NO3} and simplifying then results in equations \ref{eqn:FFR1}, \ref{eqn:FFR2}, and \ref{eqn:FFR3}.

Conversely, suppose we are given $\eta$, $f$, and $\zeta$ which solve equations \ref{eqn:FFR1}, \ref{eqn:FFR2}, and \ref{eqn:FFR3} and the fact that $\lim_{x\rightarrow \infty} \nabla \cdot \zeta = 0$. Multiply equation \ref{eqn:FFR2} by $\eta$, take the divergence of both sides, and then subtract two times equation \ref{eqn:FFR3}. Then divide by $\eta$ to get
\begin{equation}\label{eqn:DZ}
   \Delta (\nabla \cdot \zeta) - \langle \frac{\nabla \eta}{\eta}, \nabla  (\nabla \cdot \zeta)  \rangle = 0.
\end{equation}
Since $\lim_{x\rightarrow \infty} \nabla \cdot \zeta = 0$, by the maximum principle applied to this second order uniformly elliptic equation in $\nabla \cdot \zeta$, it follows that $\nabla \cdot \zeta = 0$ everywhere.

Now let $\omega = df + \zeta$. Since $\zeta$ has zero divergence, 
\begin{eqnarray}
   \Delta f &=& \nabla \cdot \omega \\
    Q_{ij}  &=&  f_{ij} + \frac12 (\zeta_{i,j} + \zeta_{j,i})= \frac12 (\omega_{i,j} + \omega_{j,i}) \\
   \Delta \zeta_i &=& \Delta (\omega_i - df_i) = \Delta\omega_i - d (\nabla \cdot \omega)_i 
\end{eqnarray}
Substituting these facts into equations \ref{eqn:FFR1} and \ref{eqn:FFR2} proves equations \ref{eqn:NO1} and
\ref{eqn:NO2}. 

\qed

By assumption, spacetime is foliated by flat three dimensional Euclidean spaces, so we have global coordinates $(t = x^0, x = x^1, y = x^2, z = x^3)$, for some interval of $t$ values.
All of the spacetime phenomena of FFR, such as gravity and the big bang, are then encoded in the real-valued function $\eta$ and the 3-vector valued function $\omega = (\omega_1, \omega_2, \omega_3)$. Said even more explicitly, the spacetime metric of the theory is determined by four real-valued functions, $\eta(t,x,y,z)$, $\omega_1(t,x,y,z)$, $\omega_2(t,x,y,z)$, and $\omega_3(t,x,y,z)$, whose values are determined by solving equations \ref{eqn:NO1} and \ref{eqn:NO2}, or equivalently equations \ref{eqn:FFR1}, \ref{eqn:FFR2}, and \ref{eqn:FFR3},
on each three dimensional Euclidean space.

Of course, there is still the issue of various choices of matter fields that one might choose to use to define the matter Lagrangian $L$. Each matter field will have its own equations of motion, determined as the Euler-Lagrange equations of $L$. These equations, which in general will depend on $\eta$ and $\omega$ as well as the matter fields, will determine how the energy density $\mu$ and momentum vector density $J$ evolve over time, and hence how $\eta$ and $\omega$ must evolve over time through equations \ref{eqn:NO1} and \ref{eqn:NO2}. Note that it therefore makes sense to pose the question of how $\eta$ and $\omega$ evolve from initial conditions of $\eta$, $\omega$, $J$, and $\mu$, even though equations \ref{eqn:NO1} and \ref{eqn:NO2} have no time derivatives.

To stay as general as possible, we will not focus on any one particular choice of matter field in this paper. 
Matter fields which come from actions which are invariant under reparametrizations of the metric (which we will always require) produce stress-energy tensors $T$ with zero divergence by Noether's theorem (first observed by Hilbert in this special case). Hence, we will assume that 
\begin{equation}
\nabla \cdot T = 0, 
\end{equation}
which may be interpreted as the local conservation of energy and momentum of the matter fields. 
Many commonly studied matter fields also satisfy the dominant energy condition on $T$, that $T(u,v) \ge 0$ for all future time-like vectors $u$ and $v$. When one of the two vectors is the normal vector to the flat foliation, the dominant energy condition 
implies
\begin{equation}\label{eqn:muJ}
   \mu \ge |J|,
\end{equation}
which corresponds to positive matter density, not exceeding the speed of light. 
For the results of this paper, we will only need to assume the weaker condition
\begin{equation}\label{eqn:muJ}
   \mu \ge 0,
\end{equation}
which corresponds to positive matter density, with no control on whether or not it exceeds the speed of light. 
Finally, it is often convenient to make assumptions on the smoothness and behavior at infinity of $\mu$ and $J$, which we will do as needed.

\section{Properties of the Flatly Foliated Relativity Equations}\label{sec:FFRprops}

One of the biggest problems for GR is that, even after a great deal of effort, no one has yet proved that it has a general, long-time existence theory. That is, given initial conditions at $t = 0$ for a spacetime, there is no guarantee that the equations of GR may be solved for all time, even in vacuum. The best guess for what one might hope to be true is known as the cosmic censorship conjecture.
However, even after more than a century since the discovery of GR, a proof of this conjecture remains elusive.

An existence theory for FFR, on the other hand, would be significantly different. While the Euler-Lagrange equations for the matter fields may still be hyperbolic, the equations which govern how the spacetime metric evolves are elliptic on each Euclidean space, as we demonstrate in this section. For example, for $J=0$ the FFR equations reduce to the $k$-Hessian equation (with $k=2$) for which there is an existence theory \cite{Hessian}.

We start by establishing that assumption \ref{axiom:arrow} yields a geometrically preferred arrow of time, which we will then adopt to prove ellipticity. Given these results, we will be in a position to pose the question of an existence theory for FFR.

\subsection{A Geometric Arrow of Time and $\Lambda > 0$}\label{sec:arrow}

In this section, we establish the intimate relationship between the requirement $\Lambda > 0$ of assumption \ref{axiom:arrow} and the existence of a geometrically preferred arrow of time. This is significant because the universe, upon inspection, does appear to have an arrow of time, and there is good evidence that the cosmological constant of the universe is greater than zero \cite{AcceleratingUniverse2}, \cite{AcceleratingUniverse1}; furthermore, the lines of reasoning leading to each of these conclusions are quite independent. Hence, the idea that these properties might in general be closely related, or in some sense even equivalent, is very interesting.

\begin{theorem} \label{thm:CC}
If $\Lambda > 0$, every $\mu \geq 0$ solution to FFR has a geometrically preferred arrow of time given by the direction of the mean curvature vector. 
\end{theorem}

{\it Proof: }  We have by equation \ref{eqn:NO1} that 
\begin{equation}
   \left(\frac{\nabla \cdot \omega}{\eta}\right)^2 \;\;\;\ge\;\;\; 2\Lambda + 16\pi\mu \;\;\;>\;\;\; 0,
\end{equation}
since $\Lambda > 0$ and $\mu \ge 0$. Hence, $\frac{\nabla \cdot \omega}{\eta}$ is never zero and so, by our smoothness assumption, can never change sign. Thus, by equation \ref{eqn:mc}, the mean curvature vector of each slice of the flat foliation is 
\begin{equation}
   \vec K = K n = \left(\frac{\nabla \cdot \omega}{\eta}\right) n \ne 0,
\end{equation}
where $n = (\partial_t + \vec\omega)/\eta$ is a smoothly defined unit normal to the flat foliation. Hence, the mean curvature vector $\vec K$, which is geometrically defined independent of any choice of coordinates or conventions, is a smooth, time-like vector, everywhere perpendicular to the foliation, with length never equal to zero, so we get an arrow of time by defining the future to be in the direction of $\vec K$.

\qed

Given assumption $\ref{axiom:arrow}$, then, the direction of $\vec K$ is either everywhere $n$ or everywhere $-n$. Without loss of generality, we will always choose our coordinate $t$ (which we could always replace with $-t$) so that the direction of $\vec K$ is everywhere $n$. 

\begin{convention}\label{convention:arrow}
When $\Lambda > 0$, we may adopt the convention that $n$ points in the direction of $\vec K$ and hence the future, from which it follows that 
\begin{equation}\label{eqn:convention}
   K = \frac{\nabla \cdot \omega}{\eta} = \frac{\Delta f}{\eta} > 0,
\end{equation}
where $f$ is defined in equation \ref{eqn:new}.
\end{convention}

Similarly, we can tell the difference between the future and the past in terms of the rate of change of the volume form of each slice. As time increases, volumes increase, when flowing between slices orthogonally.

\begin{corollary}\label{cor:VF}
When $\Lambda > 0$, flowing from slice to slice along the orthogonal flow vector $\vec\eta = \partial_t + \vec\omega$ causes the volume form of the flat foliation to increase.
\end{corollary}
{\it Proof: } The first variation formula for flowing the volume form $dV$ of each slice perpendicularly to each slice is 
\begin{equation}
   \frac{d}{dt} \; dV = - \langle \vec\eta , \vec K \rangle \; dV = \eta K \; dV.
\end{equation}

Since $\eta > 0$ and $K > 0$, the volume form is increasing. 

\qed

As an interesting remark, theorem \ref{thm:CC} has a converse: for any $\Lambda \leq 0$, there exist $\mu \geq 0$ solutions to FFR with no geometric arrow of time possible. In particular, in this case we may choose a spacetime where $16\pi\mu = -2\Lambda \ge 0$ and $J = 0$ everywhere. There exist matter fields with this property, for example one whose Lagrangian $L$ is just a constant, where $16\pi L - 2 \Lambda = 0$. The FFR equations, equations \ref{eqn:NO1} and \ref{eqn:NO2}, are then solved by letting $\eta = 1$ and $\omega = 0$, representing the Minkowski spacetime with its standard foliation of the flat, constant time slices. 

However, the Minkowski spacetime with its standard foliation is time symmetric, meaning that the operation $t \rightarrow -t$ is an isometry preserving the foliation. Hence, it is impossible for the geometry of this standard foliation on the Minkowski spacetime to produce an arrow of time; any such arrow of time would, being preserved by the $t \to -t$ isometry, be equal to its negation, a contradiction. Thus, if one fixes $\Lambda$ among the solutions to FFR, assumption \ref{axiom:arrow} is equivalent to the existence of a geometric arrow of time for every $\mu \geq 0$ solution.

\subsection{$\Lambda > 0$ Implies Ellipticity}\label{sec:elliptic}

\begin{theorem} \label{thm:E}
A positive cosmological constant implies that equations \ref{eqn:FFR1}, \ref{eqn:FFR2}, and \ref{eqn:FFR3} 
form a weakly elliptic system of equations on each flat, constant time slice. Taken one at a time on each slice with respect to the variables $f$, $\zeta$, and $\eta$, respectively, they are also each elliptic, and uniformly elliptic if there exist functions $\eta_{min}(t)$ and $Q_{max}(t)$ such that
\begin{equation}\label{eqn:precisebounds}
\eta \ge \eta_{min}(t) > 0 \;\;\;\;\;\mbox{ and }\;\;\;\;\; |Q_{ab} \nu_a \nu_b| \le Q_{max}(t), 
\end{equation}
for all unit vectors $\nu = \nu_a \partial_a$.
\end{theorem}

{\it Proof: } 
Since $\Lambda > 0$, we may use the convention in equation \ref{eqn:convention} which implies that 
$\Delta f > 0$ (where as always we continue to assume $\eta > 0$).
Hence, by equation \ref{eqn:FFR1}, for all unit vectors $\nu = \nu_a \partial_a$, where again all indices are summed from 1 to 3,
\begin{eqnarray}
\Delta f - Q_{ab} \nu_a \nu_b &=& \left( ||Q||^2 + (16\pi\mu + 2\Lambda) \; \eta^2 \right)^{1/2} - Q_{ab} \nu_a \nu_b \\
&\ge& \left( (Q_{ab} \nu_a \nu_b)^2 + 2\Lambda \; \eta^2 \right)^{1/2} - Q_{ab} \nu_a \nu_b \\
&\ge& (Q_{max}^2 + 2\Lambda \; \eta_{min}^2)^{1/2} - Q_{max} \;\;\;>\;\;\; 0,
\end{eqnarray}

since $(x^2 + c^2)^{1/2} - x$ is a decreasing function of $x$.
Note that we have not only used $\Lambda > 0$, but also that the energy density $\mu \ge 0$.

The above inequality shows that $\left( \Delta f \cdot \delta_{ab} - Q_{ab} \right)$ is uniformly positive definite on each flat, constant time slice, which is exactly what is required for the second order operator
\begin{equation}
   L = \left( \Delta f \cdot \delta_{ab} - Q_{ab} \right) \frac{\partial^2}{dx^a dx^b}
\end{equation}
to be uniformly elliptic on each slice. Ellipticity of nonlinear equations is defined in terms of their linearizations around solutions, which for equations \ref{eqn:FFR1}, \ref{eqn:FFR2}, and \ref{eqn:FFR3} are, respectively,
\begin{eqnarray}
   2 \left( \Delta f \cdot \delta_{ab} - Q_{ab} \right) &\dot{f}_{ab} & \hspace{1.65in} = \mbox{ lower order terms } \\
(\delta_{ab}) &\dot\zeta_{i,ab} & + \left( \frac{2\eta_i}{\eta} \delta_{ab} - \frac{2\eta_b}{\eta} \delta_{ia} \right) \dot{f}_{ab} = \mbox{ lower order terms } \\
   \left( \Delta f \cdot \delta_{ab} - Q_{ab} \right) & \dot{\eta}_{ab} & + \hspace{.12in} \left( \Delta \eta \cdot \delta_{ab} - \eta_{ab} \right) \hspace{.12in} \dot{f}_{ab} = \mbox{ lower order terms }
\end{eqnarray}
where we have only shown the top order variation terms, which are all that are relevant for ellipticity.
The positive definiteness of the leftmost expressions in each of the above three equations proves that 
equations \ref{eqn:FFR1}, \ref{eqn:FFR2}, and \ref{eqn:FFR3} 
are each uniformly elliptic on each slice in $f$, $\zeta$, and $\eta$, respectively. 

To show weak ellipticity for the system of equations represented by equations \ref{eqn:FFR1}, \ref{eqn:FFR2}, and \ref{eqn:FFR3}, we should first recognize that equation \ref{eqn:FFR2} is actually three equations since $i$ goes from 1 to 3, making a total of five equations. We also have five functions, which we order as $(f, \zeta_1, \zeta_2, \zeta_3, \eta)$. Keeping this ordering of the equations and the functions, the principal symbol of the system has the form
\begin{equation}
 \sigma =  \begin{bmatrix}
      + &0&0&0&0 \\
      ? & + &0&0&0 \\
      ? &0& + &0&0 \\
      ? &0&0& + &0 \\
      ? &0&0&0& + 
   \end{bmatrix}
\end{equation}
where $+$ represents a positive entry, $0$ represents a zero entry, and $?$ represents an entry whose sign is undetermined. Weak ellipticity is equivalent to this matrix always being invertible, which it is because of its lower diagonal form with positive entries along the diagonal.
\qed

Note that if $\Lambda \leq 0$, the Minkowski spacetime example given at the end of the previous section is a $\mu \geq 0$ solution wherein $\Delta f \cdot \delta_{ab} = Q_{ab} = 0$, so the conclusion of the above theorem does not hold.

\subsection{Physically Reasonable Boundary Conditions at Infinity}\label{sec:BC}

Elliptic systems of equations require boundary conditions, which in physical terms describe the asymptotics of the spacetime at infinity. Our discussion here is analogous to requiring that the Newtonian gravitational potential, for a finite amount of matter in a bounded region, go to a constant at infinity. This requirement is a boundary condition at infinity for the Poisson equation of Newtonian gravity.

Before stating the boundary conditions, it is useful to note that when $\Lambda\geq0$ there is a natural vacuum solution for $\mu=0$ and $J=0$, namely $\eta=1$, $\zeta=0$, and

\begin{eqnarray}
f = \sqrt{\frac{\Lambda}{12}}r^2 =  \sqrt{\frac{\Lambda}{12}}(x^2+y^2+z^2)
\end{eqnarray}
We now state the boundary conditions in terms of $\tilde{f} = f - \sqrt{\frac{\Lambda}{12}}r^2$, the difference from the vacuum solution. In the following, ${\bf R}^3_t$ is the flat slice of the spacetime at coordinate time $t$.

\begin{definition}
Suppose that $\Lambda > 0$, $\mu$ and $J$ are smooth, and $\mu$ and $J$ are zero outside a bounded region on ${\bf R}^3_t$, thereby representing a finite amount of matter in a bounded region, except for the positive cosmological constant. Then we will say that a solution to equations \ref{eqn:FFR1}, \ref{eqn:FFR2}, and \ref{eqn:FFR3} on ${\bf R}^3_t$ has ``physically reasonable boundary conditions at infinity'' or "physically reasonable asymptotics" if
\begin{eqnarray}
\label{eqn:BC1}
\lim_{\vec{x} \rightarrow \infty} \partial_r \tilde{f}(t,\vec{x}) &=& 0 \\
\label{eqn:BC2}
\lim_{\vec{x} \rightarrow \infty} \zeta(t,\vec{x}) &=& 0 \\
\label{eqn:BC3}
\lim_{\vec{x} \rightarrow \infty} \eta(t,\vec{x}) &=& 1
\end{eqnarray}.
\end{definition}
Note that because of its Neumann boundary condition, $\tilde{f}$ is only unique up to an additive constant, unlike the other two variables which have Dirichlet boundary conditions. One nice consequence of requiring $\lim_{\vec{x} \rightarrow \infty} \eta(t,\vec{x}) = 1$ is that this implies that the speed of the orthogonal flow of flat slices at spatial infinity is one; hence, the coordinate $t$ is the proper time experienced by observers traveling perpendicularly to the flat foliation, in the limit that they are positioned at spatial infinity. These boundary conditions are primarily motivated, however, by explicit examples of solutions we present later in the paper. These examples also suggest the following question of existence, uniqueness, and regularity of solutions to equations \ref{eqn:FFR1}, \ref{eqn:FFR2}, and \ref{eqn:FFR3}:

\begin{conjecture}\label{conjecture:existence}
Suppose that $\Lambda> 0$, $\mu$ and $J$ are smooth, and $\mu$ and $J$ are zero outside a bounded region on ${\bf R}^3_t$, thereby representing a finite amount of matter in a bounded region, except for the positive cosmological constant. Does there then exist a smooth solution $(f, \zeta, \eta)$ to equations \ref{eqn:FFR1}, \ref{eqn:FFR2}, and \ref{eqn:FFR3} on ${\bf R}^3_t$ with $\eta > 0$ and physically reasonable boundary conditions at infinity, as defined above? 
In addition, can we conclude that all of the kth partial derivatives of $d\tilde{f}$, $\zeta$, and $\eta - 1$ converge to zero at infinity as well, for $k \ge 0$? Finally, is this solution unique up to adding a constant to $\tilde{f}$?
\end{conjecture}

Since $\eta = 0$ is a solution to equation \ref{eqn:FFR3}, it is reasonable to hope that this solution may be used as a barrier, thereby guaranteeing that $\eta > 0$ for solutions with $\eta$ going to one at infinity.

Observe that, when $J=0$, the hypotheses of this question imply through equations \ref{eqn:FFR2} and \ref{eqn:FFR3} that $\zeta_i=0$ and $\eta=1$ by application of the maximum principle, justified by the previous ellipticity result. Hence, equation \ref{eqn:FFR1} reduces in this case to:
\begin{equation}
2\sum_{i\ne j}f_{ii}f_{jj} =16\pi \mu + 2\Lambda
\end{equation}
The left hand side of the above equation is the second symmetric polynomial of $\nabla^2f$, so this is the $k$-Hessian equation for $k=2$. The $J=0$ case, then, already has an existence theory akin to that suggested above \cite{Hessian}.

If the answer to the above question or one similar is yes, it would be a first step towards an existence theory for solutions to FFR coupled to matter fields, given appropriate initial conditions. 
While the question is certainly nontrivial given the full nonlinearity of equations \ref{eqn:FFR1}, \ref{eqn:FFR2}, and \ref{eqn:FFR3}, whether or not it is true should eventually be possible to determine since it is an elliptic system, as shown in the previous section. Even if the answer is yes, however, the point of highest interest is the existence theory for GR, which then leads to the following question:

\begin{conjecture}
Is there a connection between the existence theory of FFR and the existence theory of GR? More precisely, does there exist a clever foliation of spacetime (for example a scalar flat one) which simplifies the existence theory?
\end{conjecture}

\section{$\Lambda > 0$ Implies a Unique Vacuum Solution}\label{sec:vacuum}

In this section, we prove that a positive (but not a negative or zero) cosmological constant implies that there exists a unique vacuum solution to the FFR equations, equations \ref{eqn:NO1} and \ref{eqn:NO2}, among solutions which are spherically symmetric at infinity. Indeed, the FFR equations are so rigid that even mild approximate spherical symmetry is enough to deduce uniqueness of the solution when $\Lambda > 0$. There are two parts to the claim we wish to prove: existence and uniqueness. We will show that $\Lambda < 0$ fails at existence, $\Lambda = 0$ fails at uniqueness, but that $\Lambda > 0$ has both.

By contrast, GR has many vacuum solutions which are spherically symmetric at infinity, known as gravitational waves. The simplest gravitational wave solutions are perturbations of the Minkowski spacetime \cite{Wald}, though these are not spherically symmetric at infinity. Imposing the spherical symmetry at infinity condition may be achieved using results by Corvino and Schoen in \cite{corvino2006asymptotics}, at least for $\Lambda \ge 0$. The analogous $\Lambda < 0$ case would follow by a Corvino-Schoen style result in the asymptotically hyperbolic setting, which is a reasonable conjecture. 

We define ``existence'' to mean the existence of a solution $(\omega, \eta)$ which is smooth with $\eta > 0$.
Also, our notion of ``uniqueness'' considers two vacuum solutions $(\omega_1, \eta_1)$ and $(\omega_2, \eta_2)$ to be the same if they only differ by an overall positive multiplicative constant (i.e., a rescaling of the coordinate time $t$). Hence, without loss of generality, we may assume $\eta (0) = 1$, for example. Finally, we define ``spherically symmetric at infinity'' to mean that $(\omega, \eta)$ is spherically symmetric for $r \ge r_0$ in ${\bf R}^3$, for some $r_0 > 0$.

The purpose of the ``spherically symmetric at infinity'' hypothesis is two-fold. First, this hypothesis highlights a difference between GR and FFR. Second, without this hypothesis, the next theorem would not be true. A nice way of understanding the necessity of this hypothesis is to return to the Poisson equation $\Delta V = 4\pi \mu = 0$ in the vacuum case. If  the Newtonian potential $V$ is required to go to zero at infinity, then the maximum principle implies that $V = 0$ everywhere. However, without any control at infinity, $V$ could be any harmonic function, such as $2x^2 - y^2 - z^2$, which one could think of as representing infinite amounts of matter, infinitely far away. Requiring spherical symmetry at infinity is a condition which rules out these exotic solutions. In particular, spherical symmetry at infinity means that on a sufficiently large sphere, $V = c$ for some constant $c$, which by the maximum principle implies that $V = c$ everywhere.

In the case of $\Lambda>0$, instead of proving directly the existence and uniqueness of vacuum solutions which are spherically symmetric at infinity, we prove a slightly stronger result:

\begin{theorem} \label{thm:UV}
Assume $\Lambda>0$ and $f$ is approximately spherically symmetric at $\infty$,  whereby we mean $f-\bar{f}$ is bounded and
$$
\lim_{r\to\infty} \sup_{x\in S_r} r^2 \|\nabla^2(f-\bar f)\nabla(f-\bar f)\| \to 0,
$$
where $\bar{f}$ is the spherical symmetrization of $f$. Then under the boundary conditions (\ref{eqn:BC2}) and (\ref{eqn:BC3}),  there is a unique (up to an additive constant on $f$) smooth vacuum solution to the FFR equations in elliptic form, equations (\ref{eqn:FFR1})-(\ref{eqn:FFR3}).
\end{theorem}

{\it Proof: } In the vacuum case, the elliptic FFR equations reduce to
\begin{eqnarray}
\label{eqn:FFR1V} (\Delta f)^2 - ||Q_{ij}||^2 &=& 2\Lambda \; \eta^2, \\
\label{eqn:FFR2V} \Delta \zeta_i &=& - \frac{2}{\eta}\left( \Delta f \cdot \delta_{ij} - Q_{ij} \right) \eta_j, \\
\label{eqn:FFR3V} \left( \Delta f \cdot \delta_{ij} - Q_{ij} \right) \eta_{ij} &=& 0.
\end{eqnarray}
Observe that by the boundary condition $\eta \to 1$ at $\infty$, for every $\epsilon>0$ there exists an $r_0 > 0$ such that for any $r \ge r_0$, we have 
$$
1-\epsilon <\eta<1+\epsilon
$$
on $S_r$.
Since $\Lambda>0$, theorem \ref{thm:E} yields that equation (\ref{eqn:FFR3V}) is elliptic, so we may apply the maximum principle to $\eta$ on $B_r$, which yields that the above bounds hold on all of $B_r$. Since this is true for every $r \geq r_0$, the bounds hold on all of $\mathbb{R}^3$. Since $\epsilon > 0$ was arbitrary, we conclude $\eta \equiv 1$ and thus $\Delta \zeta_i=0$.

Similarly, applying the maximum principle to $\zeta_i$, $1\le i\le 3$, with the boundary condition $\zeta \to 0$ at $\infty$, we have $\zeta \equiv 0$.

Hence, we have reduced the theorem to proving there exists a unique solution to
\begin{equation}
   (\Delta f)^2 - ||f_{ij}||^2 = 2\Lambda
\end{equation}
which is approximately spherically symmetric at infinity. Following our standard arrow of time convention (this has been implicitly invoked already by citing theorem \ref{thm:E}'s ellipticity result), $\Delta f > 0$, so this is equivalent to
\begin{equation}
   \Delta f = \sqrt{2\Lambda + ||f_{ij}||^2}.
\end{equation}

Recall that the spherical symmetrization of a function $\psi(r,\theta,\phi)$ in ${\bf R}^3$ is defined to be
\begin{eqnarray}
   \bar\psi (r) &=& \frac{1}{|S^2|} \int_{S^2} \psi(r,\theta,\phi) \; d\sigma_{S^2} \\
                   &=& \frac{1}{4\pi} \int_0^\pi \int_0^{2\pi} \psi(r,\theta,\phi) \; \sin\theta \, d\phi \, d\theta.
\end{eqnarray}
Equivalently, $\bar\psi$ is the average over all (origin fixing) rotations $R \in SO(3)$ of $\psi$,
\begin{eqnarray}
   \bar\psi ({x}) &=& \frac{1}{|SO(3)|} \int_{R \in SO(3)} \psi(R(x)) \; d\sigma_{SO(3)} ,
\end{eqnarray}
where we have abused notation slightly with $\bar\psi ({x}) = \bar\psi (|{x}|) = \bar\psi (r)$. 
We have
\begin{align*}
&\frac{1}{|SO(3)|}\int_{R\in SO(3)}\| f_{ij}(R(x))\|d\sigma_{SO(3)}
\\\ge&\frac{1}{|SO(3)|}\left\|\int_{R\in SO(3)}f_{ij}(R(x))d\sigma_{SO(3)}\right\|
\end{align*}
by Jensen's inequality with equality if and only if $\|\cdot\|$ is linear.
However, this is by the triangle inequality only the case if $f_{ij}(R(x))$ are all contained in the same 1-dimensional subspace.
Thus equality implies that $f_{ij}(R(x))=f_{ij}(x)$ or $f_{ij}(R(x))=-f_{ij}(x)$ and hence by continuity $f=\Bar f$.

Using once more Jensen's inequality applied to the convex function $y = \sqrt{2\Lambda + x^2}$, we obtain
\begin{equation}
   \Delta \bar{f} = \overline{\Delta f} = \overline{\sqrt{2\Lambda + ||f_{ij}||^2}} 
                      \ge \sqrt{2\Lambda + \overline{||f_{ij}||}^2}
                      \ge \sqrt{2\Lambda + ||\bar{f}_{ij}||^2},
\end{equation}
so that  
\begin{equation}
   (\Delta \bar{f})^2 - ||\bar{f}_{ij}||^2 \ge 2\Lambda,
\end{equation}
with equality if and only if the Hessian of $f(R(x))$ is equal to the Hessian of $\bar{f}(x)$, for all rotations $R \in SO(3)$, which is the case if and only if $f = \bar{f}$ (since two function with equal Hessians differ by a linear function, and $f - \bar{f}$ is bounded). In summary, we have:

\begin{lemma}
Given a function $f$ which is approximately spherically symmetric at infinity and satisfies
\begin{equation}
   (\Delta f)^2 - ||f_{ij}||^2 = 2\Lambda,
\end{equation}
its spherical symmetrization $\bar{f}$ satisfies
\begin{equation}\label{eqn:FE}
   (\Delta \bar{f})^2 - ||\bar{f}_{ij}||^2 \ge 2\Lambda,
\end{equation}
with equality if and only if $f$ is spherically symmetric everywhere.
\end{lemma}

Our next step to prove theorem \ref{thm:UV} is based on the identity 
\begin{eqnarray}
   \nabla \cdot \left( \Delta f \, \nabla f - \Hess f (\nabla f, \cdot) -\frac{\Lambda}{3} \cdot 2r \partial_r \right) 
   &=& (f_{ii}f_j - f_{ij}f_i)_j - \frac{\Lambda}{3} \Delta (r^2)
   \\&=& (\Delta f)^2 - ||f_{ij}||^2 - 2\Lambda,
\end{eqnarray}
since $\Delta (r^2) = \Delta (x^2 + y^2 + z^2) = 6$. 

Using the assumption that $f$ is approximately spherically symmetric at $\infty$, we obtain for any $\delta>0$, there exists an $r_1>0$ such that for any $R\ge r_1$, we have:
\begin{eqnarray}
&\,&\int_{B_R} \left[(\Delta \bar{f})^2 - ||\bar{f}_{ij}||^2 - 2\Lambda \right] \; dV
\\&=& \int_{S_R} \left\langle \Delta \bar{f} \, \nabla \bar{f} - \Hess \bar{f} (\nabla \bar{f}, \cdot) -\frac{\Lambda}{3} \cdot 2r \partial_r , \partial_r \right\rangle \; dA
\\&=&\int_{S_R} \left\langle \Delta f
\nabla f - \Hess f (\nabla f, \cdot) -\frac{\Lambda}{3} \cdot 2r \partial_r , \partial_r \right\rangle \; dA
\\&\,&-\int_{S_R} \left\langle \Delta(\bar{f}-f)\nabla(\bar{f}-f)-\Hess(\bar{f}-f)(\nabla(\bar{f}-f),\cdot), \partial _r\right\rangle \; dA \label{eqn:deltaBound}
\\&\,&+\int_{S_R} \left\langle \Delta(\Bar{f}-f)\nabla\bar{f}+\Delta\bar{f}\nabla(\bar{f}-f),\partial_r \right\rangle \; dA \label{eqn:vanish1}
\\&\,&-\int_{S_R} \left\langle \Hess(\Bar{f}-f)(\nabla\bar{f},\cdot)+\Hess\bar{f}(\nabla(\bar{f}-f),\cdot),\partial_r \right\rangle \; dA \label{eqn:vanish2}
\end{eqnarray}
Hereby line \eqref{eqn:deltaBound} can be bounded by $\delta$ by approximate spherical symmetry.
The first term of line \eqref{eqn:vanish1} vanishes since 
\begin{eqnarray}
&\,&\int_{S_R} \left\langle \Delta(\Bar{f}-f)\nabla\bar{f},\partial_r \right\rangle \; dA 
\\&=&\partial_r \Bar f\int_{S_R}(\Delta_{S_R}+\frac{2}R\partial_r+\partial_{r,r})(\Bar f-f)\; dA
\\&=&\partial_r\Bar f(\frac{2}R\partial_r+\partial_{r,r})\int_{S_R}\Bar f-f\; dA=0.
\end{eqnarray}
The second term vanishes in a similar fashion.
For line \eqref{eqn:vanish2}, observe
$$
 \Hess(\Bar{f}-f)(\nabla\bar{f},\partial_r)=\partial_{r,r}(\Bar f-f)\partial_r\Bar f
$$
and
$$
\Hess\bar{f}(\nabla(\bar{f}-f),\partial_r)=\partial_{r,r}\Bar f\partial_r(\Bar f-f).
$$
Since $\partial_r \Bar f$, $\partial^2_{r,r}\Bar f$ are constant on $S_R$ and we can move $\partial^2_{r,r}$, $\partial_r$ outside the integral, line \eqref{eqn:vanish2} vanishes too.
Hence we obtain
\begin{eqnarray}
&\,&\int_{B_R} \left[(\Delta \bar{f})^2 - ||\bar{f}_{ij}||^2 - 2\Lambda \right] \; dV
\\&\le&  \int_{S_R} \left\langle \Delta f
\nabla f - \Hess f (\nabla f, \cdot) -\frac{\Lambda}{3} \cdot 2r \partial_r , \partial_r \right\rangle \; dA+\delta
\\&=& \int_{B_R} \left[(\Delta {f})^2 - ||{f}_{ij}||^2 - 2\Lambda \right] \; dV+\delta
= \delta.
\end{eqnarray}

By the lemma, the original integrand is always nonnegative, so the integral increases with $R$; thus, $\delta$ is a bound on the original integral for every $R>0$. Since $\delta>0$ was arbitrary, the integral vanishes for each $R>0$, showing $(\Delta \bar{f})^2 - ||\bar{f}_{ij}||^2 - 2\Lambda \equiv 0$ by smoothness. By the lemma again, then, $f$ is spherically symmetric.

A standard computation yields that the eigenvalues of the Hessian of a spherically symmetric function $f(r)$ are $f_{rr}$, $f_r/r$ and $f_r/r$ again. Hence, 
\begin{eqnarray}
2\Lambda &=& (\Delta f)^2 - ||f_{ij}||^2 \label{eqn:SScalc1}
\\&=& 
\left(f_{rr} + 2 \, \frac{f_r}{r} \right)^2 - \left( f_{rr}^2 + 2 \, \frac{ f_r^2}{r^2}  \right) \\
&=& 
\frac{4}{r} \, f_r \, f_{rr} + \frac{2}{r^2} f_r^2 
\\&=&
\frac{2}{r^2} \left( r f_r^2 \right)_r \label{eqn:SScalc2}
\end{eqnarray}
which implies that 
\begin{equation}\label{eqn:UV}
  f = \sqrt{\frac{\Lambda}{12}} \, r^2 + c,
\end{equation}
for some constant $c$, since $f_r = 0$ at $r = 0$ for $f$ to be smooth. This proves that $f$ is unique up to the constant $c$ and completes the proof of theorem \ref{thm:UV}.

 \qed

We now use this theorem to deduce, in the $\Lambda > 0$ case, the existence and uniqueness of a vacuum solution to FFR that is spherically symmetric at infinity.

\begin{corollary}\label{cor:UV}
Suppose the cosmological constant $\Lambda > 0$. Then there exists a unique vacuum solution to the FFR equations, equations \ref{eqn:NO1} and \ref{eqn:NO2}, among solutions which are spherically symmetric at infinity.
\end{corollary}

 {\it Proof:} For existence, take the solution ($f,\zeta,\eta$) to the elliptic equations produced in theorem \ref{thm:UV} and consider that $\zeta \equiv 0$, so $\nabla \cdot \zeta \equiv 0$ and theorem \ref{thm:equivalence} implies that this gives rise to a vacuum solution ($\omega,\eta$), with $\omega = df$, to the FFR equations, equations \ref{eqn:NO1} and \ref{eqn:NO2}. This solution is spherically symmetric.

For uniqueness, we show that any vacuum solution ($\omega,\eta$) to FFR that is spherically symmetric at infinity induces a smooth vacuum solution to the FFR equations in elliptic form satisfying the hypotheses of theorem \ref{thm:UV}. Let $\omega = \omega_{SS} + \omega_{C}$, where $\omega_{SS}=\omega_{SS}(r)dr$ is spherically symmetric and $\omega_{C}$ has compact support, which is possible since $\omega$ is spherically symmetric outside a finite radius. 
Then we may define $f = f_{SS} + f_{C}$, where $d f_{SS} = \omega_{SS}$ and $\Delta f_{C} = \nabla \cdot \omega_{C}$ are solved, respectively, by integrating $f_{SS}'(r) = \omega_{SS}(r)$ and by convolving with the Green's function $\Phi$ for the Laplacian in $\mathbb{R}^3$. By theorem \ref{thm:equivalence}, the triple ($f,\zeta,\eta$) with $\zeta = \omega - df = \omega_C - df_C$, which satisfies $\nabla \cdot \zeta = 0$, is a vacuum solution to the elliptic FFR equations, equations \ref{eqn:FFR1}, \ref{eqn:FFR2}, and \ref{eqn:FFR3}.

By spherical symmetry at infinity and the maximum principle, $\eta$ is constant, so since our notion of uniqueness of ($\omega,\eta$) is equality up to an overall positive multiplicative constant, we may take $\eta \equiv 1$ (recall that $\eta$ is restricted to $\eta > 0$), and in particular boundary condition (\ref{eqn:BC3}) is satisfied. We claim that $\zeta = \omega-df$ is identically zero; since $\eta$ is constant, we have that $\Delta \zeta_i =0$ by equation \ref{eqn:FFR2}, so it suffices (as in the proof of theorem \ref{thm:UV}) to show $\zeta \to 0$ at infinity, or that boundary condition (\ref{eqn:BC2}) is satisfied. Outside the support of $\omega_C$, we have $\zeta = -df_C$, where as noted before
\begin{equation}
f_C(x) = \int_{\mathbb{R}^3} \Phi(y) (\nabla \cdot \omega_C)(x-y) dy.
\end{equation} 
Since $\omega_C$ is smooth and has compact support, standard analysis arguments yield that 
\begin{equation}
\nabla f_C(x) = \int_{\mathbb{R}^3} \Phi(y) (\nabla (\nabla \cdot \omega_C))(x-y) dy = \int_{\mathbb{R}^3} \Phi(x-y) (\nabla (\nabla \cdot \omega_C))(y) dy,
\end{equation} 
so $\|\zeta\| = \|df_C \| = \|\nabla f_C\| \to 0$ as $\|x\| \to \infty$ since, in the last integral above, $\Phi(x-y)$ decays like $\|x-y\|^{-1}$ and $y$ can be restricted to the support of $\omega_C$. Thus, $\zeta \equiv 0$.

Finally, this implies that $\omega = df$, so $f$ is spherically symmetric at infinity because $\omega$ is, and hence $f$ is approximately spherically symmetric at infinity. By theorem \ref{thm:UV}, then, $f$ is of the form given in equation \ref{eqn:UV}, showing $\omega = df = \sqrt{\frac{\Lambda}{3}}rdr$, establishing uniqueness. 

\qed

{\bf Theorem \ref{thm:intro} has now been proved}: it follows from corollaries \ref{cor:VF} and \ref{cor:UV} and theorems \ref{thm:CC} and \ref{thm:E}.
Now that we have proved that $\Lambda > 0$ implies that there exists a unique vacuum solution to the FFR equations with spherical symmetry at infinity, we will show that  $\Lambda = 0$ fails at uniqueness and $\Lambda < 0$ fails at existence.

\begin{lemma}\label{lem:UV-zero}
Suppose the cosmological constant $\Lambda = 0$. Then there exist many vacuum solutions to the FFR equations, equations \ref{eqn:NO1} and \ref{eqn:NO2}, among solutions which are spherically symmetric at infinity.
\end{lemma}
{\it Proof: }
By theorem \ref{thm:equivalence}, we obtain a solution to the FFR equations, equations \ref{eqn:NO1} and \ref{eqn:NO2}, from each solution to the FFR equations in elliptic form, equations \ref{eqn:FFR1}, 
\ref{eqn:FFR2}, and \ref{eqn:FFR3}, which for $\mu = 0$, $J = 0$, and $\Lambda = 0$ are
\begin{eqnarray}
\label{eqn:FFR1-0} (\Delta f)^2 - ||Q_{ij}||^2 &=& 0 \\
\label{eqn:FFR2-0} \Delta \zeta_i &=& - \frac{2}{\eta}\left( \Delta f \cdot \delta_{ij} - Q_{ij} \right) \eta_j \\
\label{eqn:FFR3-0} \left( \Delta f \cdot \delta_{ij} - Q_{ij} \right) \eta_{ij} &=& 0
\end{eqnarray}
\noindent
where $Q_{ij} = f_{ij} + \frac12 (\zeta_{i,j} + \zeta_{j,i})$ and we require that $\lim_{x\rightarrow \infty} \nabla \cdot \zeta = 0$. Further, any ($f,\zeta,\eta$) of the form
\begin{eqnarray}
   f &=& 0 \\
   \zeta &=& 0 \\
   \eta &=& \mbox{\it any function which is spherically symmetric at infinity}
\end{eqnarray}
is a vacuum solution to the above system which is spherically symmetric at infinity.

 \qed

Clearly, the proof of the above lemma even gives many examples of solutions which satisfy the physically reasonable boundary conditions defined in section $\ref{sec:BC}$.

\begin{lemma}\label{lem:UV-negative}
Suppose the cosmological constant $\Lambda < 0$. Then there does not exist a vacuum solution to the FFR equations, equations \ref{eqn:NO1} and \ref{eqn:NO2}, among solutions which are spherically symmetric at infinity.
\end{lemma}

{\it Proof: }
Suppose there exists a vacuum solution which is spherically symmetric at infinity. 
We will now derive a contradiction, proving that such a solution is not possible.

As in the proof of corollary \ref{cor:UV}, given such a solution ($\omega,\eta$) to the FFR equations, we may write $\omega = \omega_C + \omega_{SS}$ where $\omega_C$ has compact support and $\omega_{SS}$ is spherically symmetric, and further we may find a spherically symmetric function $f_{SS}$ such that $\omega_{SS} = df_{SS}$, so that $\omega = df_{SS}$ outside the support of $\omega_C$. Now writing $f=f_{SS}$, we have that, as in the derivation of the FFR equations in elliptic form, substituting  $\omega =df$ in the FFR equations with $\mu=0$, $J=0$ gives
\begin{eqnarray}
\label{eqn:FFR1-neg} (\Delta f)^2 - ||f_{ij}||^2 =& 2\Lambda \; \eta^2 \\
\label{eqn:FFR2-neg} \left( \Delta f \cdot \delta_{ij} - f_{ij} \right) \eta_j =& 0
\end{eqnarray}
\noindent
outside the support of $\omega_C$, say for $r \geq r_0$; we now restrict our considerations to this region.
Simplifying equation \ref{eqn:FFR1-neg} as done in equations \ref{eqn:SScalc1} to \ref{eqn:SScalc2}, we get
\begin{equation}\label{eqn:125}
 \left( r f_r^2 \right)_r = \Lambda r^2 \eta^2 < 0
\end{equation}
which shows that $r f_r^2$ is a strictly decreasing function of $r$ since $\Lambda < 0$ and smooth solutions require $\eta > 0$. Since it is true in general that $r f_r^2 \ge 0$, this further shows that $r f_r^2$ can never equal zero. Thus, 
\begin{equation}
   f_r \ne 0.
\end{equation}
On the other hand, equation \ref{eqn:FFR2-neg} (which is really three equations, one for each $1 \leq i \leq 3$) implies that 
\begin{equation}
   0 = \left[\left(f_{rr} + \frac{2}{r} f_r \right) - f_{rr} \right] \eta_r = \frac{2}{r} f_r \eta_r
\end{equation}
so that
\begin{equation}
   \eta_r = 0
\end{equation}
and hence $\eta = c$, for some constant $c>0$ on our region of interest. Inserting this back into equation \ref{eqn:125} we get
\begin{equation}
 \left( r f_r^2 \right)_r = \Lambda c^2 r^2 < 0,
\end{equation}
which is clearly not possible since the integral of the right hand side will eventually make $r f_r^2$ negative, a contradiction. Hence, when $\Lambda < 0$, there does not exist a vacuum solution which is spherically symmetric at infinity. 

\qed

\section{Explicit Solutions to Flatly Foliated Relativity}

Now that we have established some of the basic theory of FFR, we present some of the most important explicit solutions of the theory. After we reproduce the results of GR for cosmological spacetimes describing the big bang and the accelerating expansion of the universe, we show how the Schwarzschild spacetime metric is a distributional solution to the FFR equations, where a finite amount of matter is concentrated at the origin as a Dirac-delta distribution on each flat Euclidean slice.

\subsection{Friedmann–Lemaître–Robertson–Walker Spacetimes}

FFR is consistent with the standard FLRW cosmological spacetimes of GR, but with the requirement that the curvature of each constant time slice is $k = 0$. These spacetimes, along with the fact that observations are in agreement with the hypothesis $k = 0$ \cite{planck}, are one of the main motivations for FFR.

The usual form of these spacetime metrics, in spherical coordinates, is 
\begin{eqnarray}
   g &=& -dt^2 + a(t)^2 \delta_{{\bf R}^3} \\
      &=& -dt^2 + a(t)^2 \left( dR^2 + R^2 d\sigma^2 \right),
\end{eqnarray}
where $d\sigma^2 = d\theta^2 + \sin(\theta)^2 d\phi^2$ is the metric on the unit sphere. As usual, $a(t)$ represents the relative linear size of the universe as a function of time.
To convert the above form to the standard form for FFR as in equation \ref{eqn:Metric}, we define a new coordinate $r = a(t) R $, which implies
\begin{eqnarray}
   R &=& \frac{r}{a(t)} \\
 a(t) dR &=& dr - H(t) \; r \; dt,
\end{eqnarray}
where 
\begin{equation}
H(t) = \frac{a'(t)}{a(t)}
\end{equation}
is the ``Hubble constant,'' which is actually a function of time. 
Hence, in $(t,r)$ coordinates, the spacetime metric is 
\begin{equation}
   g = \left( -1 + H(t)^2 r^2 \right) dt^2 - H(t) \, r \, (dr \otimes dt + dt \otimes dr) + dr^2 + r^2  d\sigma^2 ,
\end{equation}
where we note that the last two terms are the flat metric on ${\bf R}^3$ in spherical coordinates. Hence, this spacetime metric is in the form of equation \ref{eqn:Metric} with $\eta = 1$ and $\omega = H(t) \, r \, dr$. Equivalently, in our elliptic coordinates, 
\begin{eqnarray}
   f &=& \frac12 \, H(t) \, r^2 = \frac12 \, H(t) \, (x^2 + y^2 + z^2)\\
   \zeta &=& 0 \\
   \eta &=& 1 .
\end{eqnarray}
Plugging this into equations \ref{eqn:FFR2} and \ref{eqn:FFR3} just gives $J = 0$, as is true in the FLRW models. Since 
\begin{eqnarray}
   f_{ij} &=& H(t) \, \delta_{ij} \\
   \Delta f &=& 3 H(t),
\end{eqnarray}
equation \ref{eqn:FFR1} implies that 
\begin{eqnarray}
16\pi \mu + 2 \Lambda &=& (\Delta f)^2 - ||f_{ij}||^2 \\
                                 &=& 9 H(t)^2 - 3 H(t)^2 ,
\end{eqnarray}
which yields
%
\begin{center}
{\bf The First Friedmann Equation} 
\begin{equation}\label{eqn:Friedmann1}
   \frac{8\pi \mu + \Lambda}{3} = H(t)^2,
\end{equation}
\end{center}
with $k = 0$. Note that $\mu \ge 0$ and $\Lambda > 0$ implies that $H(t) \ne 0$, so that once the universe starts expanding, it can never stop. The above equation is also the statement that, in this model with $k=0$, the density of the universe is always at the critical density.

FLRW spacetimes are assumed to be filled with matter whose stress energy tensor is homogeneous and isotropic on each constant time slice. The implication of this assumption is that the stress energy tensor for each matter field has the form
\begin{equation}\label{eqn:SET}
T = \begin{bmatrix}
      \mu(t) & 0 & 0 & 0 \\
          0 & P(t) & 0 & 0 \\
          0 & 0 & P(t) & 0 \\
          0 & 0 & 0 & P(t)
      \end{bmatrix} 
= - \mu g + (\mu + P) \delta_{{\bf R}^3},
\end{equation}
where the matrix represents the components of $T$ with respect to the orthonormal frame $\{n, \partial_1, \partial_2, \partial_3\}$. As usual, we assume $\nabla \cdot T = 0$, and in lemma \ref{lem:dot} we will prove the well known result, but using our language, that this implies 
\begin{center}
{\bf The Conservation Equation} 
\begin{equation}\label{eqn:conservation}
   \mu'(t) = -3 H(t) (\mu + P) 
\end{equation}
\end{center}
for each matter field. 

Together, equations \ref{eqn:Friedmann1} and \ref{eqn:conservation} form a system of ordinary differential equations whose solutions very accurately model the observed expansion of the universe \cite{AcceleratingUniverse2}, \cite{AcceleratingUniverse1}. 
 Equation \ref{eqn:Friedmann1} determines the rate of change of the relative size of the universe as represented by $a(t)$ and equation \ref{eqn:conservation} determines the rate of change of the matter density of each matter field (typically by assuming that for each individual matter field, $P = w \mu$, for some constant $w$ determined by the nature of the matter field). 
Thus, FFR derives the same equations and hence gives exactly the same results as GR for FLRW spacetimes with $k = 0$.

We comment that equations \ref{eqn:Friedmann1} and \ref{eqn:conservation} also imply the second Friedmann equation, for which we have no need.
It is worth noting that we can not derive the second Friedmann equation the usual way by taking the trace of the Einstein equation $G = 8\pi T - \Lambda g$. This is because, as described in section \ref{sec:FFR}, in FFR we only have equations \ref{eqn:Enn} and \ref{eqn:Enp}. However, we still get all that we need in this case from the fact that the divergence of the stress energy tensor is still zero. This suggests  $\nabla \cdot T = 0$ is very important for FFR in other settings as well, especially when we want to understand how matter is conserved.

Equation \ref{eqn:conservation} also implies conservation laws for matter fields depending on how their pressure relates to their energy density. Let $P = w \mu$, for some $-1 \le w \le 1$ (so as not to violate the dominant energy condition). For example, the cosmological constant (a.k.a. dark energy) has $w = -1$, cold dark matter and regular baryonic matter moving at slow speeds are approximated with $w =0$, and electromagnetic radiation is approximated by $w = 1/3$, as explained in standard texts \cite{Wald}. Then 
\begin{equation}
   \mu(t) \cdot a(t)^{3(1+w)} = \mbox{constant}.
\end{equation}
The time derivative of the above conservation identity is equivalent to equation \ref{eqn:conservation}.

\begin{lemma}\label{lem:dot}
$\nabla \cdot T = 0$ implies equation \ref{eqn:conservation}.
\end{lemma}
{\it Proof: }
Since the covariant derivative of a metric is always zero, the hardest part of taking the divergence of equation \ref{eqn:SET} is computing the divergence of $\delta = \delta_{{\bf R}^3}$. Note that 
\begin{eqnarray}
\left(\nabla \cdot \delta \right) (n) 
&=& \sum_{\alpha = 0}^3 \left( \nabla_{\partial_\alpha} \delta \right) (\partial_\alpha, n) 
= \sum_{\alpha = 0}^3 \partial_\alpha \left( \delta (\partial_\alpha, n) \right) - \delta (\nabla_{\partial_\alpha}\partial_\alpha, n) - \delta (\partial_\alpha, \nabla_{\partial_\alpha}n) 
\nonumber \\&=& - \sum_{j = 1}^3 \delta (\partial_j, \nabla_{\partial_j}n) 
= - \sum_{j = 1}^3 g (\partial_j, \nabla_{\partial_j}n)
= \sum_{j = 1}^3 g (\nabla_{\partial_j}\partial_j, n)
\nonumber \\&=& - \tr k = - \Delta f = - 3 H(t).
\end{eqnarray}
Hence, 
\begin{eqnarray}
0 &=& \left(\nabla \cdot T \right) (n) 
\nonumber \\&=& -d\mu(n) + (\mu + P) \left(\nabla \cdot \delta \right) (n)
\nonumber \\&=& - \mu'(t) + (\mu + P) \left( - 3 H(t) \right),
\end{eqnarray}
which proves equation \ref{eqn:conservation}.

\qed

\subsection{The Schwarzschild Solution}\label{sec:Schwarzschild}

The Schwarzschild solution to GR may, in a precise sense, \textit{almost} be given a flat foliation via the Gullstand-Painlevé coordinates to make it a solution to FFR. While this spacetime has $\Lambda = 0$ and hence does not satisfy the hypotheses invoked in section \ref{sec:FFRprops}, we will see that the mean curvature vector associated to the foliation (or equivalently, $\nabla \cdot \omega = \Delta f$) is everywhere nonzero, so we may still adopt convention \ref{convention:arrow}. Further, the Schwarzschild spacetime with this foliation and convention is a limit of qualitatively similar spacetimes that are honest $\Lambda > 0$ solutions to FFR, so we understand the properties of these by studying the Schwarzschild case.

Understanding the Schwarzschild solution in FFR is very instructive in several regards. For instance, the FFR solution for the Schwarzschild spacetime never enters the black hole and so avoids the black hole singularity. In other words, FFR does not attempt to solve for the spacetime metric in the vicinity of the Schwarzschild black hole singularity. This suggests the possibility that the FFR equations might have a nice existence theory.

While the flat foliation does enter the white hole and intersects the white hole singularity, FFR treats this singularity as a point particle with a Dirac-delta matter distribution at the origin of each flat three dimensional Euclidean space. In fact, while curvatures blow up at the white hole singularity, the FFR equations are still solvable at the singularity in the distributional sense. In addition, one may perturb away the Schwarzschild singularity entirely, if one wishes, within the class of spherically symmetric spacetimes by convolving the matter density with a bump function. In doing so, the topology changes: the punctured three dimensional flat slices of the Schwarzschild spacetime become complete three dimensional Euclidean spaces. 

Although we are about to show how a Schwarzschild white hole can be thought of as a point particle with a Dirac-delta matter distribution at the origin, this does not mean that this represents a physical matter field satisfying the dominant energy condition.
Since inside a white hole $r$ is always increasing along future timelike curves,
physical matter fields focused at a point would be expected to be unstable and fly apart very quickly.

This interpretation of the Schwarzschild white hole spacetime as a static point particle raises the question of what a moving point particle would look like. It also suggests the possibility that point particles, complete with their own dynamics, might be a rigorous concept which exists in FFR, completely analogous to the Newtonian $n$-body problem. If so, this could be a way to approximately model stars, planets, and moons, and perhaps even binary pulsars.

The first part of this section presents the Schwarzschild spacetime in three coordinate systems: Kruskal coordinates, static coordinates, and then Gullstrand–Painlevé coordinates. The Gullstrand–Painlevé coordinates are a special case of FFR coordinates, equation \ref{eqn:Metric}, when solving the FFR equations, equations \ref{eqn:FFR1}, \ref{eqn:FFR2}, and \ref{eqn:FFR3}, in the distributional sense with $\mu = m \cdot \delta_0$, $J = 0$, and $\Lambda = 0$,  where $\delta_0$ is the Dirac-delta distribution with integral one, located at the origin. We then show how to understand this solution as the limit of solutions with smooth matter densities $\mu$ which are concentrating at the origin, and how the solution changes with a positive cosmological constant.

\subsubsection{Coordinate Chart Representations of the Schwarzschild Spacetime}\label{sec:SchwarzschildCoords}

\begin{figure}
\begin{center}
\includegraphics[height = 3.1in]{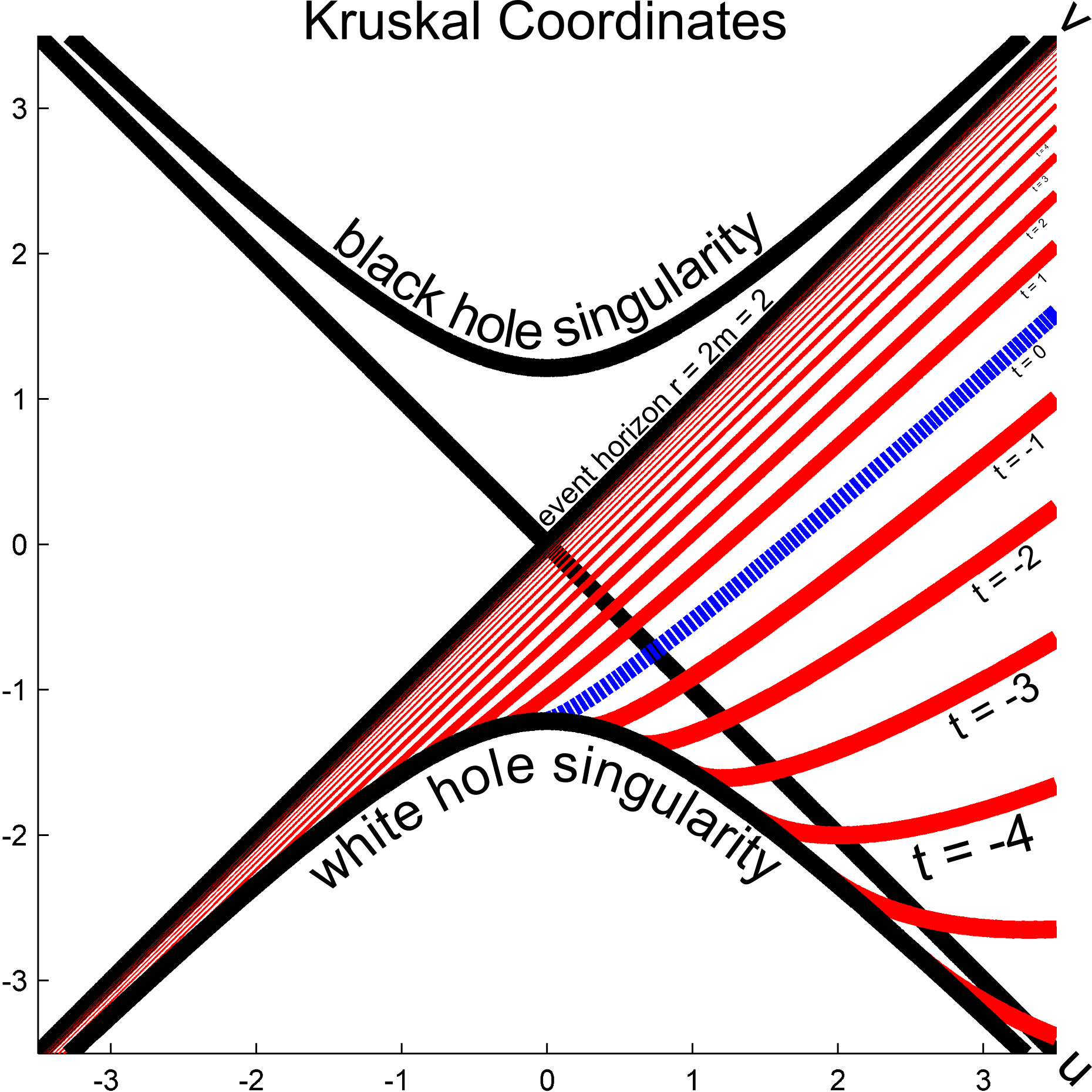} \hspace{.2in}
\includegraphics[height = 3.1in]{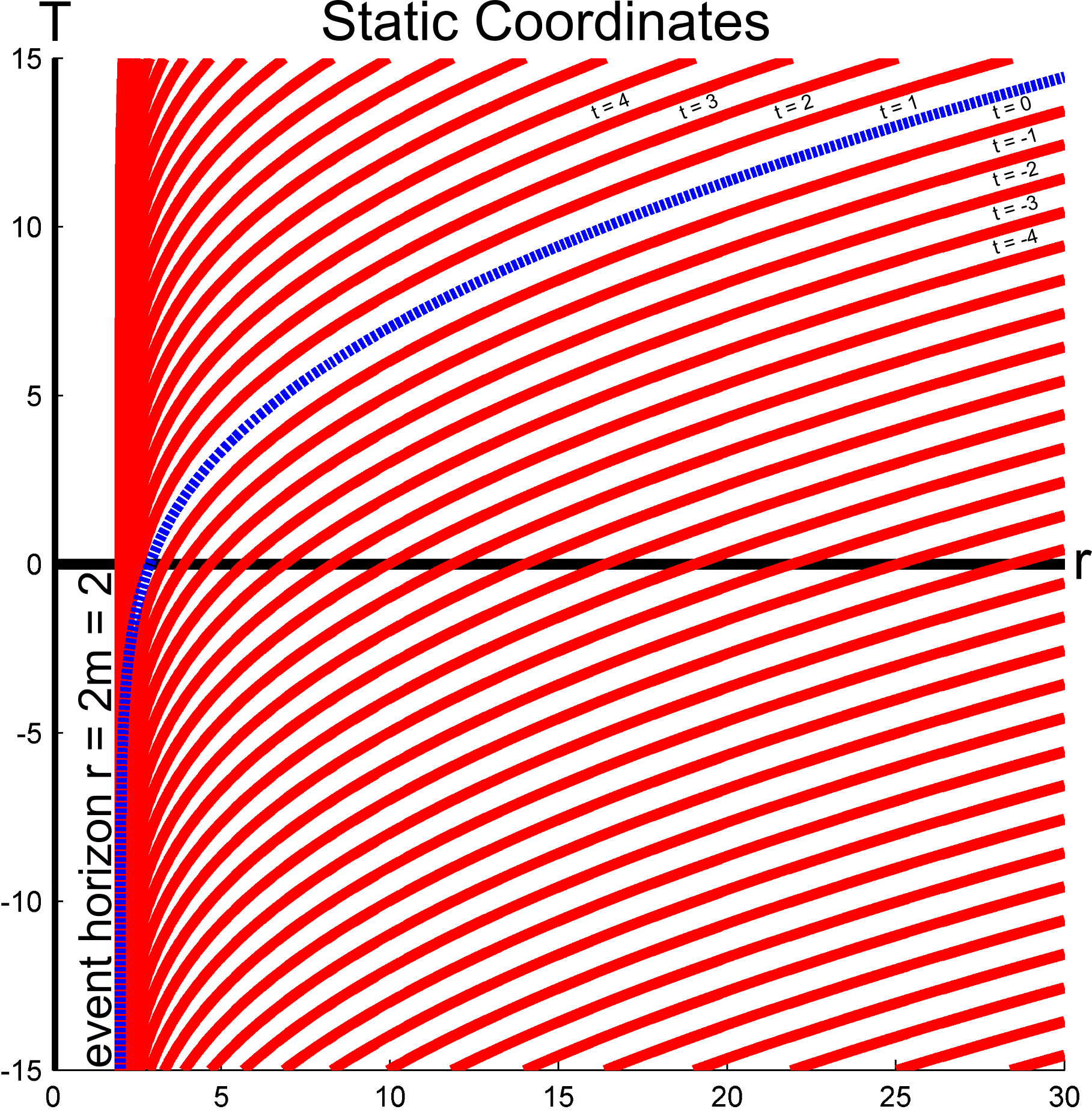}
\end{center}
\caption{\label{fig:Schwarzschild} The images above show how the unit mass Schwarzschild spacetime (Kruskal coordinates on the left and static coordinates on the right) may be foliated by flat three dimensional Euclidean spaces. In both cases, our arrow of time, defined by the mean curvature vector, points up, the orthogonal speed of the foliation $\eta = 1$, and $-\infty < t < \infty$, with the integer valued constant $t$ slices drawn and the $t=0$ slice colored blue. Note that the foliation on the left, while it enters the white hole, never enters the black hole, even though it continues to flow into the future, perpendicularly to each slice, with speed one.}
\end{figure}

The three most important coordinate chart representations of the Schwarzschild spacetime, for our purposes, are listed below:

\begin{center}
{\bf Kruskal Coordinates} 
\begin{equation}
g = F(r) \left( du \otimes dv + dv \otimes du \right) + r^2 d\sigma^2
\end{equation}
\end{center}
\vspace{.04in}
\begin{center}
{\bf Static Coordinates} 
\begin{equation}
g = - \left( 1 - \frac{2m}{r} \right) dT^2 + \left( 1 - \frac{2m}{r} \right)^{-1} dr^2 + r^2 d\sigma^2
\end{equation}
\end{center}
\vspace{.04in}
\begin{center}
{\bf Gullstrand–Painlevé Coordinates for Flatly Foliated Relativity} 
\begin{equation}\label{eqn:gschwarz}
g = - \left( 1 - \frac{2m}{r} \right) dt^2 - \sqrt{\frac{2m}{r}} \left( dt \otimes dr + dr \otimes dt \right) + dr^2 + r^2 d\sigma^2
\end{equation}
\end{center}
where  
\begin{eqnarray}
F(r) &=& \frac{8m^2}{r} \exp{\left(1-\frac{r}{2m}\right)} , \\
uv \;\;=\;\; f(r) &=& (r - 2m) \exp \left( \frac{r}{2m} - 1 \right) , \\
\mbox{and }\hspace{.35in} d\sigma^2 &=& d\theta^2 + \sin(\theta)^2 d\phi^2
\end{eqnarray}
is the metric on the unit sphere of radius one, in spherical coordinates. Technically, we have replaced $t$ with $-t$ in the usual form of the Gullstrand–Painlevé coordinates to be compatible with our arrow of time that requires the mean curvature vector of the flat slices to point in the direction of the future. The effect of this time reversal is that, whereas the usual Gullstrand–Painlevé coordinates foliate the exterior region of Schwarzschild and the inside of the black hole, our coordinates foliate the exterior region of Schwarzschild and the inside of the white hole, as depicted in figure \ref{fig:Schwarzschild}.

With our conventions, 
Kruskal coordinates are $(u,v,\theta,\phi)$ (where $r = f^{-1}(uv)$), static coordinates are 
$(T,r,\theta,\phi)$, and Gullstrand–Painlevé coordinates are $(t,r,\theta,\phi)$. Note that there is some overlap in these coordinates. For example, all three sets of coordinates use the same angle coordinates $(\theta,\phi)$ and any spherically symmetric sphere with radius coordinate $r$ is always defined to have area $4\pi r^2$. Finally, the isometric transformations between these coordinate charts is summarized by the equations
\begin{eqnarray}
\sqrt{f(r)} \exp \left(- \frac{T}{4m} \right) = &u& = \left( \sqrt{r} + \sqrt{2m} \right) 
\exp \left( \frac{r-t}{4m} -\sqrt{\frac{r}{2m}} - \frac12 \right) \\
\sqrt{f(r)} \exp \left( \frac{T}{4m} \right) = &v& = \left( \sqrt{r} - \sqrt{2m} \right) 
\exp \left( \frac{r+t}{4m} +\sqrt{\frac{r}{2m}} - \frac12 \right) 
\end{eqnarray}
which imply that
\begin{equation}
   T = 2m \left( \log\left( \sqrt{r} - \sqrt{2m} \right) - \log\left( \sqrt{r} + \sqrt{2m} \right) \right)
      + \sqrt{8mr} + t .
\end{equation}
These relationships are depicted graphically in figure \ref{fig:Schwarzschild}.

\subsubsection{The Schwarzschild Spacetime as a Distributional Solution to the FFR Equations}
\label{sec:SchwarzschildSolution}

Note that the above Gullstrand–Painlevé coordinates fit the form of our flatly foliated spacetimes in equation \ref{eqn:Metric} with $\eta = 1$ and $\omega = \sqrt{\frac{2m}{r}} \, dr$, which is equivalent to 
\begin{eqnarray}
f &=& 2 \sqrt{2mr} \\
\zeta &=& 0 \\
\eta &=& 1.
\end{eqnarray}
The above values for $f$, $\zeta$, and $\eta$ are, as one would expect, a solution to 
equations \ref{eqn:FFR1}, 
\ref{eqn:FFR2}, and \ref{eqn:FFR3} with $\mu = 0$, $J = 0$, and $\Lambda = 0$, away from the singularity at $r = 0$, since the Schwarzschild spacetime is vacuum with zero cosmological constant. However, if we include the singularity, we get something importantly different. 

\begin{theorem}
The above values for $f$, $\zeta$, and $\eta$ are a solution in the distributional sense to equations \ref{eqn:FFR1}, 
\ref{eqn:FFR2}, and \ref{eqn:FFR3}, where 
\begin{eqnarray}
   \mu &=& m \cdot \delta_0 \\
       J &=& 0,
\end{eqnarray}
$\Lambda = 0$, and $\delta_0$ is the Dirac delta distribution with integral one, located at the origin. 
\end{theorem}
{\it Comment: }
Note that while Gullstrand–Painlevé coordinates require $r \ne 0$, the FFR equations have no such restriction. The equations are true in the distributional sense for all $r$. Hence, we have effectively filled in the puncture $r \ne 0$, thereby changing the topology of each constant time slice to that of ${\bf R}^3$. Even so, this is not strictly a solution to FFR in the sense defined in section \ref{sec:FFR}, as the associated metric cannot be smoothly extended to $r=0$.

{\it Proof: } Since $\zeta$ = 0, $\eta = 1$, and $J = 0$, equations \ref{eqn:FFR2} and \ref{eqn:FFR3} are automatically satisfied. Hence, all we need to do is to verify equation \ref{eqn:FFR1}, which in this case is the distributional equation
\begin{equation}\label{eqn:dis}
 (\Delta f)^2 - ||f_{ij}||^2 = 16\pi  m \cdot \delta_0 . 
\end{equation}
Let $\phi$ be any smooth test function with compact support. Then
\begin{eqnarray}
\int_{{\bf R}^3} \phi \left[ (\Delta f)^2 - ||f_{ij}||^2 \right] \; dV
&=&
\int_{{\bf R}^3} \phi \;\; \nabla \cdot \left( \Delta f \, df -  \Hess f (\nabla f, \cdot) \right) \; dV
\label{eqn:l1}\\\label{eqn:l2} &=& 
- \int_{{\bf R}^3} \left\langle d\phi \; , \; \Delta f \, df -  \Hess f (\nabla f, \cdot) \right\rangle \; dV
\\ &=&
- \int_{{\bf R}^3} \left\langle d\phi \; , (f_{rr} + \frac2r f_r) f_r \, dr - f_{rr} \, f_r \, dr \right\rangle \; dV
\\ &=&
- \int_0^\infty \left\langle d\bar\phi \; , \frac2r f_r^2  dr \right\rangle \; 4\pi r^2 dr
\\ &=&
- \int_0^\infty \bar\phi_r \cdot 16\pi m \; dr
\\ &=& 
16\pi m \cdot \bar\phi(0) 
\\ &=& 
16\pi m \cdot \phi(0) 
\end{eqnarray}
where $\bar\phi$ is the spherical symmetrization of $\phi$ around the origin. This proves equation \ref{eqn:dis}.

\qed

{\it Comment: } Since the FFR equations (equations \ref{eqn:FFR1}, \ref{eqn:FFR2}, and \ref{eqn:FFR3}) may be written in divergence form (equations \ref{eqn:FFR1D}, \ref{eqn:FFR2D}, and \ref{eqn:FFR3D}), weak solutions in the distributional sense to these equations may be defined. To be more precise, the integrals in equations \ref{eqn:l1} are {\it defined} to equal the integral in equation \ref{eqn:l2}. A weak solution, then, is any solution to the equations which holds true when integrated against any smooth test function with compact support, as above.

\subsubsection{ The Spherically Symmetric, $J=0$ Case}

Another way to understand the Schwarzschild spacetime as a distributional solution to the elliptic FFR equations is to exhibit it as a limit of smooth solutions. To that end, here we find all smooth $J=0$ solutions to FFR in the spherically symmetric case when the cosmological constant $\Lambda > 0$. Recall that in the spherically symmetric case, we may always solve $\omega =df$, so it suffices to consider the elliptic FFR equations, equations \ref{eqn:FFR1}, \ref{eqn:FFR2}, and \ref{eqn:FFR3}, with $J=0$, $\zeta \equiv 0$. As we've seen before, spherical symmetry and ellipticity of the last equation imply that $\eta$ is constant, so without loss of generality we may rescale the coordinate time $t$ such that $\eta \equiv 1$. Hence, we are left with only the first equation, which becomes

\begin{equation}
 (\Delta f)^2 - ||f_{ij}||^2 = 16\pi \mu + 2\Lambda . 
\end{equation}
Reducing the left hand side of this result as in equations \ref{eqn:SScalc1} to \ref{eqn:SScalc2}, we have
\begin{equation}
\frac{2}{r^2} \left( r f_r^2 \right)_r = 16\pi \mu + 2\Lambda .
\end{equation}
It then follows that 
\begin{equation}
   r f_r^2 = 2M(r) + \frac{\Lambda}{3} r^3
\end{equation}
where by symmetry and smoothness we have imposed $f_r(0) = 0$ and where
\begin{equation}
   M(r) = \int_0^r \mu(\rho) \cdot 4\pi \rho^2 \, d\rho  ,
\end{equation}
which we recognize as the integral of the energy density $\mu$ inside the sphere of radius $r$. Hence, it follows that, up to an irrelevant integration constant, 
\begin{equation}\label{eqn:fint}
   f(r) = \int_0^r \left( \frac{\Lambda}{3} \rho^2 + \frac{2M(\rho)}{\rho} \right)^{1/2} d\rho .
\end{equation}

Now suppose that the cosmological constant is zero (the above is still a solution in this case, though we no longer expect it is unique), and further that $\mu(r) \ge 0$ is smooth and $M(r) = m$ for $r > \delta$, for some constant $m$. Then in the limit as $\delta$ goes to zero, equation \ref{eqn:fint} implies that $f(r)$ converges to $2 \sqrt{2mr}$, the Schwarzschild white hole solution. Hence, the Schwarzschild spacetime may be realized as the limit of smooth FFR spacetimes with zero cosmological constant and $J = 0$, in the limit as the matter density $\mu(r)$ concentrates at the origin.
 
More generally, from equations \ref{eqn:Metric} and \ref{eqn:fint} we conclude that the general form of the metric for a spherically symmetric, $J=0$ FFR solution is
\begin{equation}\label{eqn:ggen}
g = - \left( 1 - \frac{\Lambda}{3} r^2 - \frac{2M(r)}{r} \right) dt^2 - \sqrt{\frac{\Lambda}{3} r^2 + \frac{2M(r)}{r}} \left( dt \otimes dr + dr \otimes dt \right) + dr^2 + r^2 d\sigma^2 .
\end{equation}
Compare this result to the Schwarzschild case of equation \ref{eqn:gschwarz}. Setting $M(r) = m$ in this equation then gives the analog to the Schwarzschild solution with cosmological constant, sometimes called the deSitter Schwarzschild spacetime (the above metric with $M(r)=m$ is expressed in slightly different coordinates than the standard presentation of the deSitter Schwarzschild metric-- the required change is analogous to that from static coordinates to Gullstrand-Painlevé coordinates seen in section \ref{sec:SchwarzschildCoords}). If the restriction of $\mu$ to each Euclidean slice is compactly supported, equation \ref{eqn:ggen} agrees with the deSitter Schwarzschild solution outside the support of $\mu$ and tends towards the unique vacuum solution of section \ref{sec:vacuum} as $r \to \infty$. One can make this more general by allowing $\mu$ to be a function of time as well, in which case $M(r)$ simply becomes $M(r,t)$ in the obvious way.

\bibliographystyle{plain}
\bibliography{references.bib}

\

\

\textsc{Hubert Bray}, \texttt{hubert.bray@duke.edu}

\textsc{Benjamin Hamm}, \texttt{benjamin.hamm@duke.edu}

\textsc{Sven Hirsch}, \texttt{sven.hirsch@duke.edu}

\textsc{James Wheeler}, \texttt{james.c.wheeler@duke.edu}

\textsc{Yiyue Zhang}, \texttt{yiyue.zhang@duke.edu}

\

\noindent
\textsc{Department of Mathematics and Physics, Duke University,  Durham, NC 27708-0320}

\end{document}